\documentclass[sigconf]{acmart}
\AtBeginDocument{%
  }

\copyrightyear{2026}
\acmYear{2026}
\setcopyright{cc}
\setcctype{by}
\acmConference[MM '26]{Proceedings of the 34th ACM International Conference on Multimedia}{November 10--14, 2026}{Rio de Janeiro, Brazil}
\acmBooktitle{Proceedings of the 34th ACM International Conference on Multimedia (MM '26), November 10--14, 2026, Rio de Janeiro, Brazil}
\acmDOI{10.1145/3767308.3835962}
\acmISBN{979-8-4007-2213-4/2026/11}

\acmSubmissionID{5841}

\usepackage{pifont}    
\usepackage{makecell}  
\usepackage{multirow}
\usepackage{balance}

\usepackage{xcolor}
\usepackage[framemethod=TikZ]{mdframed}
\usepackage{listings}
\usepackage{booktabs}
\usepackage{array}
\usepackage{algorithm}
\usepackage{algorithmic}
\usepackage{amsmath}
\usepackage{mathtools}
\usepackage{amsthm}
\usepackage{xurl}

\usepackage{float}

\definecolor{boxframe}{RGB}{43,82,152}
\definecolor{boxheader}{RGB}{210,230,255}
\definecolor{codeblue}{HTML}{0000FF}
\definecolor{codegreen}{HTML}{098658}
\definecolor{codemaroon}{HTML}{A31515}

\mdfdefinestyle{myfaststyle}{
    linecolor=boxframe,
    linewidth=0.8pt,
    backgroundcolor=white,
    frametitlerule=true,
    frametitlerulewidth=0.5pt,
    frametitlerulecolor=boxframe,
    frametitlebackgroundcolor=boxheader,
    frametitlefont=\bfseries\large\color{black},
    roundcorner=3mm,
    innertopmargin=10pt,
    innerbottommargin=10pt,
    innerleftmargin=10pt,
    innerrightmargin=10pt,
    skipabove=10pt,
    skipbelow=10pt,
    splittopskip=12pt
}

\newmdenv[style=myfaststyle]{promptbox_inner}

\newenvironment{promptbox}[1]
  {
    \vspace{10pt}
    \begin{promptbox_inner}[frametitle={#1}]
    \lstset{
        basicstyle=\small\ttfamily,
        breaklines=true,
        breakatwhitespace=false,
        columns=fullflexible,
        keepspaces=true,
        frame=none,
        keywordstyle=\color{black},
        commentstyle=\color{black},
        stringstyle=\color{black}
    }
  }
  {
    \end{promptbox_inner}
  }

\lstnewenvironment{codebox}[2][]
  {
    \mdfsetup{
        style=myfaststyle,
        frametitle={#2}
    }
    \mdframed
    \lstset{#1}
  }
  {
    \endmdframed
  }

\usepackage{etoolbox}

\newcommand{\publishedpages}{10}

\makeatletter

\patchcmd{\@mkbibcitation}
  {\ref{TotPages}~\@pages@word}
  {\publishedpages~pages}
  {}{}

\patchcmd{\@mkbibcitation}
  {\ref{TotPages}~\@pages@word}
  {\publishedpages~pages}
  {}{}

\makeatother

\begin{document}

\title{DesignAsCode: Bridging Structural Editability and Visual Fidelity in Graphic Design Generation}
\thanks{This arXiv version extends the 10-page paper accepted to ACM Multimedia 2026 by including supplementary material.}

\author{Ziyuan Liu}
\authornote{Work done during the authors’ internship at Microsoft Research Asia.}
\affiliation{%
  \department{School of Software and Microelectronics}
  \institution{Peking University}
  \city{Beijing}
  \country{China}
}
\email{liuziyuan@stu.pku.edu.cn}

\author{Shizhao Sun}
\authornote{Corresponding author.}
\affiliation{%
  \institution{Microsoft}
  \city{Beijing}
  \country{China}}
\email{shizhaosun@outlook.com}

\author{Danqing Huang}
\affiliation{%
  \institution{Microsoft}
  \city{Mountain View}
  \country{USA}
}

\author{Yingdong Shi}
\authornotemark[1]
\affiliation{%
 \institution{ShanghaiTech University}
 \city{Shanghai}
 \country{China}}

\author{Meisheng Zhang}
\authornotemark[1]
\affiliation{%
  \department{School of Software and Microelectronics}
  \institution{Peking University}
  \city{Beijing}
  \country{China}}

\author{Ji Li}
\affiliation{%
  \institution{Microsoft}
  \city{Mountain View}
  \country{USA}}

\author{Jingsong Yu}
\affiliation{%
  \institution{Peking University}
  \city{Beijing}
  \country{China}}

\author{Jiang Bian}
\affiliation{%
  \institution{Microsoft}
  \city{Beijing}
  \country{China}}

\renewcommand{\shortauthors}{Ziyuan Liu et al.}

\begin{abstract}
Graphic design generation demands a delicate balance between high visual fidelity and fine-grained structural editability. However, existing approaches typically bifurcate into either non-editable raster image synthesis or abstract layout generation devoid of visual content. Recent combinations of these two approaches attempt to bridge this gap but often suffer from rigid composition schemas and unresolvable visual dissonances (e.g., text-background conflicts) due to their inexpressive representation and open-loop nature. To address these challenges, we propose DesignAsCode, a novel framework that reimagines graphic design as a programmatic synthesis task using HTML/CSS. Specifically, we introduce a Plan-Implement-Reflect pipeline, incorporating a Semantic Planner to construct dynamic, variable-depth element hierarchies and a Visual-Aware Reflection mechanism that optimizes the code to rectify rendering artifacts. Extensive experiments demonstrate that DesignAsCode significantly outperforms baselines in both structural validity and aesthetic quality. Furthermore, our code-native representation unlocks advanced capabilities, including automatic layout retargeting, complex document generation (e.g., resumes), and CSS-based animation. Our project page is available at \url{https://liuziyuan1109.github.io/design-as-code/}.
\end{abstract}



\begin{teaserfigure}
  \includegraphics[
    width=\textwidth,
    trim=0 250 0 10,
    clip
]{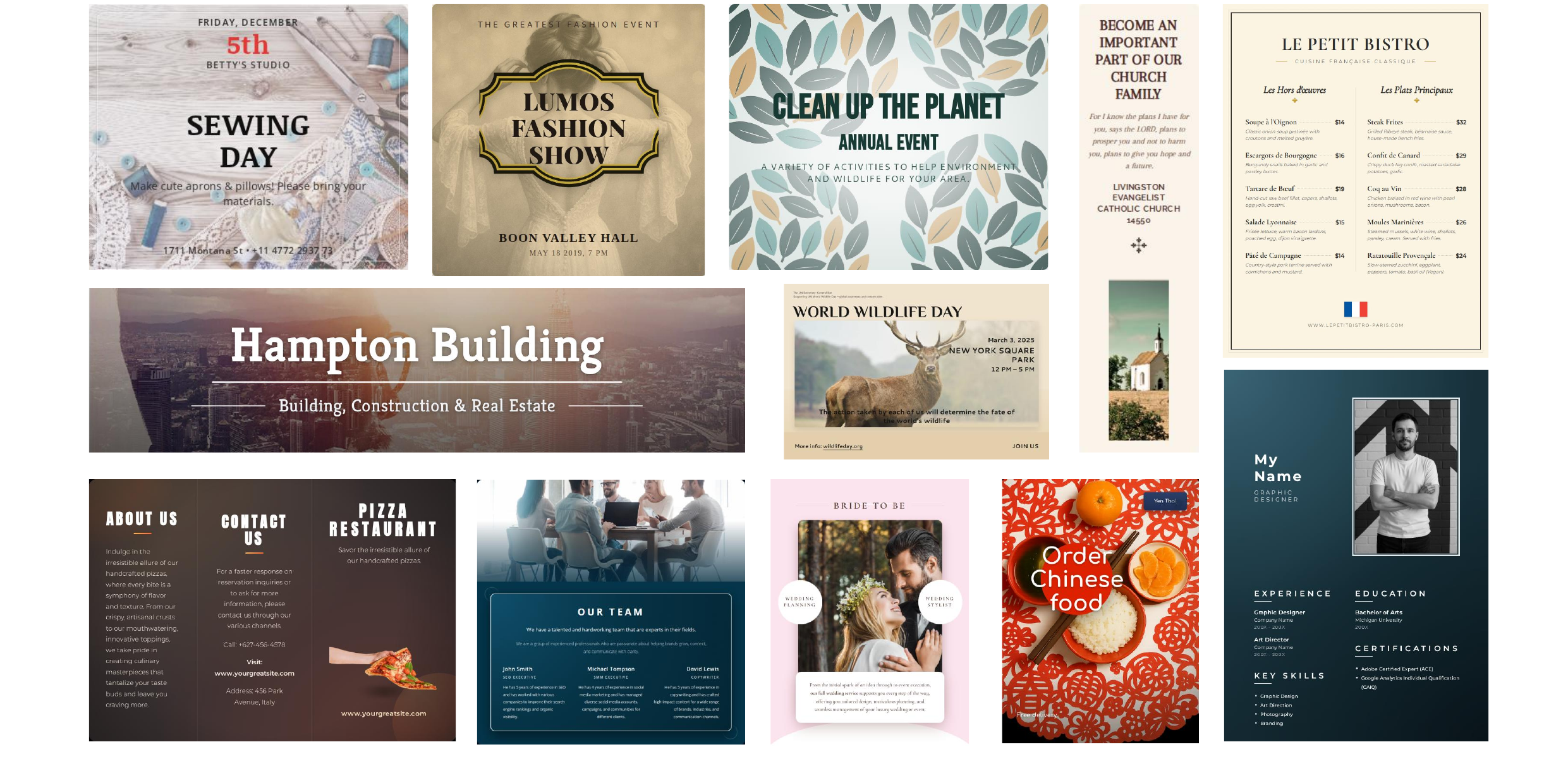}
  \caption{\textbf{Generation results from DesignAsCode.} Our method achieves superior visual fidelity across various design types (e.g., banner, flyer, menu, and resume). Furthermore, by utilizing HTML/CSS as the backend representation, it supports rich visual effects and ensures full editability for image, text, layout, color, and font attributes.}
  \Description{This image displays a grid of twelve diverse graphic design layouts, such as promotional posters, menus, and resumes, showcasing complex and highly coordinated arrangements of typography, background imagery, and decorative elements.}
  \label{fig:teaser}
\end{teaserfigure}

\settopmatter{authorsperrow=4}
\maketitle

\section{Introduction}
Graphic design is a ubiquitous and highly specialized creative activity encompassing a wide array of formats, including advertising posters, book covers, social media banners, and marketing brochures. Unlike general artistic creation, professional graphic design is strictly constrained by functional requirements: it must convey specific information while maintaining aesthetic appeal. In this context, editability is paramount. Professional workflows are rarely one-shot; they require adjustments where elements must remain distinct and manipulable. A non-editable design, no matter how visually striking, is effectively useless in a production pipeline that demands precise typographic control and flexible layout modification.

Early attempts to automate this process largely bifurcated into two disconnected streams. On one hand, large-scale \textbf{Text-to-Image (T2I) models}~\cite{rombach2022high, podell2023sdxl, betker2023improving}, despite achieving photorealistic fidelity, generate ``flat" raster images where text and background are inextricably fused, rendering them unsuitable for design modification, \textbf{lacking structure editability}. On the other hand, \textbf{layout generation} works~\cite{gupta2021layouttransformer, lin2023layoutprompter, hsu2023posterlayout} focused solely on predicting bounding boxes for potential elements. While structurally editable, these methods output abstract placeholders devoid of pixel-level detail, inherently \textbf{lacking visual fidelity}. Consequently, they leave the burden of visual realization entirely to users, who must manually source and prepare design assets, risking stylistic inconsistency and aesthetic fragmentation when the collected elements fail to harmonize.

Recently, there has been growing effort on combining layout generation model with Text-to-Image model~\cite{cole, inoue2024opencole, zhang2025creatiposter, chen2025posta}. The core philosophy of these relevant works is to train a specialized layout model to predict attributes (e.g., position, scale, font size) which then serve as a skeleton for element combination. In this paradigm, visual assets (images) are typically handled via a modular approach: they are either pre-generated to condition the layout model or generated post-hoc to fill the predicted slots. This ``divide-and-conquer" strategy attempts to emulate the layer-wise construction of a design file.
However, these attempts to bridge the gap between structural editability and visual fidelity via \textbf{naive combinations of T2I and layout generation models} remain far from satisfactory, failing to produce professional-grade editable designs. These methods have three main problems: 1) \textbf{Rigid and Inexpressive Representation:} Current methods typically rely on rigid, primitive representations (e.g., basic JSON or coordinate lists). This constrains the design space, preventing the generation of rich stylistic attributes common in professional design, such as backdrop filters, gradient transitions, or complex layer blending modes. 2) \textbf{Fixed-Schema Composition:} Most existing works operate under a constrained ``slot-filling" paradigm, typically restricted to a fixed hierarchy of one background and one foreground. This rigid composition capability lacks the flexibility to model complex, variable layouts that require dynamic numbers of image layers, nested groupings, or irregular visual hierarchies. Such fixed schemas prevent the generation of diverse and structural rich designs, forcing the output into repetitive, template-like patterns that fail to adapt to complicated content requirements. 3) \textbf{Unresolvable Visual Dissonance:} Prevailing frameworks largely adopt an open-loop, cascaded workflow. This unidirectional process prevents deep interaction between structure and content, frequently resulting in irreversible visual conflicts. For instance, a layout model might blindly position white text over a region that the image generator subsequently renders as a high-luminance highlight. Without a visual feedback mechanism to backtrack and rectify these artifacts, the system produces designs with compromised readability and broken visual hierarchies, rendering the output unusable.

To this end, we propose \textbf{DesignAsCode}, a novel framework that reimagines graphic design as a programmatic synthesis task, grounded in executable HTML/CSS code and governed by a ``Plan-Implement-Reflect" pipeline. \textbf{First,} we establish \textbf{executable HTML/\allowbreak CSS} as the unified representation space for design generation. Unlike rigid, primitive representations that constrain stylistic depth, our code-based approach natively supports rich rendering attributes---such as \textit{backdrop-filter}, \textit{mix-blend-mode}, and complex gradients. This allows the model to generate professional-grade visual effects that were previously unattainable in simple stacking of layers. This choice also aligns naturally with the structural priors of modern LLMs, as HTML/CSS is prevalent in web-scale pretraining corpora, thereby better leveraging their pretrained capabilities.
\textbf{Second,} we introduce a \textbf{Semantic Planner} to jointly reason over content synthesis and structural arrangement by distilling compositional priors from professional design corpora. In contrast to existing works constrained by fixed-schema compositions (i.e., the ``background-foreground" slot-filling), our planner dynamically constructs complex, variable-depth element hierarchies based on content semantics. Subsequently, an Implementation module translates this abstract plan into a preliminary executable HTML/CSS draft, populating the hierarchy with generated assets to produce a renderable prototype. This flexibility enables the generation of sophisticated layouts involving irregular groupings and adaptive layer counts, overcoming the structural monotony of template-based approaches. \textbf{Third,} we devise a \textbf{Visual-Aware Reflection} mechanism to refine the prototype. Addressing the structural-visual disconnect inherent in open-loop pipelines, this module acts as a specialized ``Art Director," detecting and rectifying visual dissonances (e.g., text-background color conflicts or occlusion). Specifically, it renders the preliminary code into an image and employs a vision model to propose pixel-level rectifications. This optimized visual result then serves as a reference target to guide the corresponding updates in the HTML/CSS code. This ensures that the final output maintains both high readability and aesthetic harmony. Overall, our contributions are summarized as follows:

\begin{itemize}
    \item We pioneer the \textbf{Design-as-Code paradigm} via HTML/CSS synthesis. This approach transcends primitive representations to enable professional-grade styling, native editability, and direct browser execution.
    
    \item We propose a \textbf{Plan-Implement-Reflect pipeline} integrating semantic planning and visual-aware reflection. This framework enables reasoning over complex hierarchies and autonomously rectifies rendering dissonances to ensure structural and aesthetic harmony.
    
    \item Experiments demonstrate that DesignAsCode \textbf{significantly outperforms baselines}. Furthermore, our code-native representation unlocks advanced capabilities like automatic layout retargeting, complex document generation, and CSS animation.
\end{itemize}

\begin{figure}[t!]
  \centering
  \includegraphics[
    width=\columnwidth,
    trim=0 440 0 0,
    clip
  ]{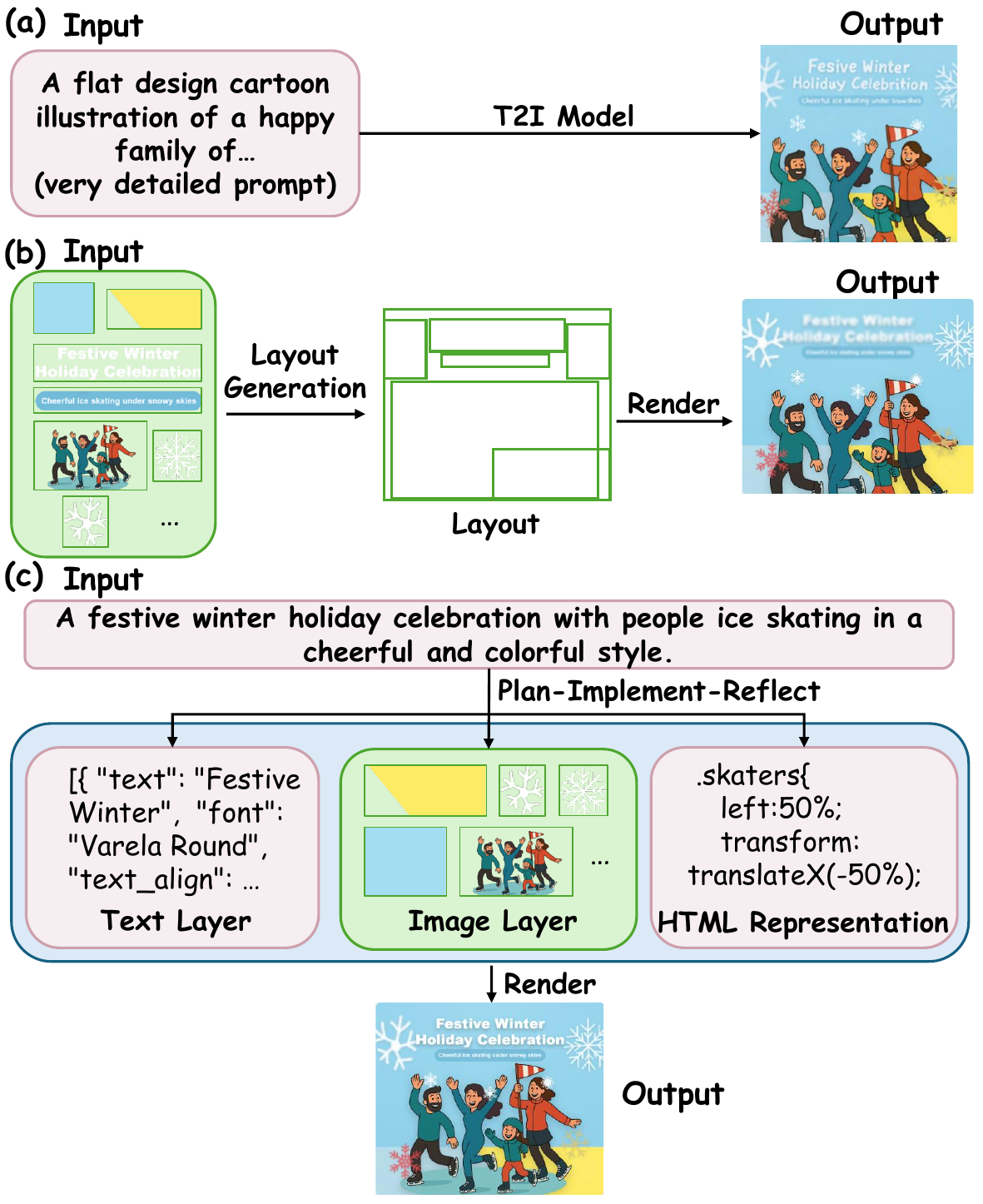}
  \caption{\textbf{Comparison of three mainstream approaches.} (a) Text-to-Image. It Lacks editability. (b) Layout generation. While creating editable design, it suffers from poor visual harmony and put heavy burden on human to prepare design assets. (c) Our editable design generation. It formulates graphic design as a programmatic synthesis using HTML/CSS, thus achieving structural editability and high visual fidelity at the same time.}
  \Description{This flowchart visually compares three graphic design generation pipelines: direct text-to-image synthesis, wireframe layout generation with placeholder rendering, and the proposed "Plan-Implement-Reflect" method that uniquely generates distinct text layers, image layers, and HTML code to render a fully structured final output.}
  \label{fig:3_main_approaches}
\end{figure}

\begin{figure*}[t]
  \centering
  \includegraphics[
    width=\textwidth,
    trim=20 10 20 25,
    clip
  ]{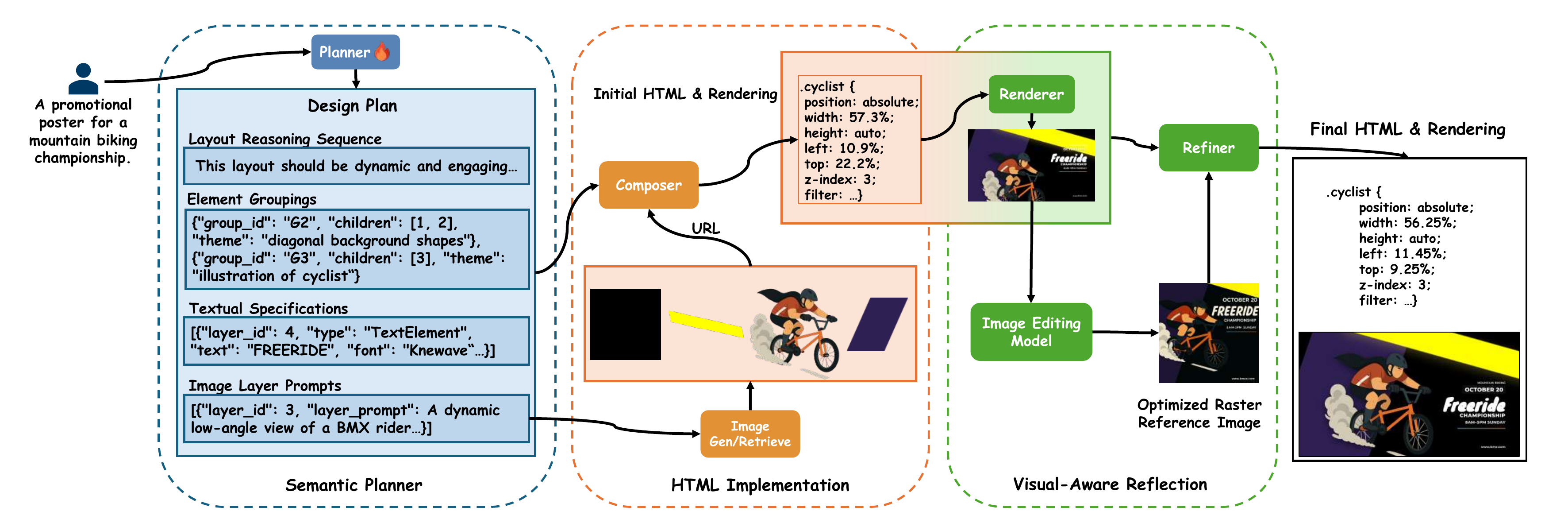}
  \caption{\textbf{Overall framework of our approach.}}
  \Description{This system architecture diagram outlines a three-stage pipeline—Semantic Planner, HTML Implementation, and Visual-Aware Reflection—illustrating how a text prompt is progressively transformed into a structured design plan, drafted into initial HTML code with generated image assets, and optimized into a final layout using visual feedback.}
  \label{fig:pipeline}
\end{figure*}

\section{Related Works}

\begin{table}[t]
\caption{\textbf{Comparison of Capabilities in Text-to-Editable-Design Methods.} }
\label{tab:comparison}
\begin{center}
\begin{small}
\begin{sc}
\setlength{\tabcolsep}{6pt} 
\begin{tabular}{l c c c}
\toprule
Method & \makecell[c]{Expressive \\ Representation} & \makecell[c]{Flexible \\ Schema} & \makecell[c]{Visual \\ Coherence} \\
\midrule
OpenCOLE      & \ding{55}    & \ding{55}    & \ding{55} \\
POSTA         & \ding{55}    & \ding{55}    & \ding{55} \\
IGD           & \ding{55}    & \ding{51}& \ding{55} \\
CreatiPoster  & \ding{55}    & \ding{55}    & \ding{55} \\
PosterCopilot & \ding{55}    & \ding{55}    & \ding{51} \\
BannerAgency  & \ding{55}    & \ding{55}    & \ding{51} \\
PosterVerse   & \ding{51}    & \ding{55}    & \ding{55} \\
\textbf{DesignAsCode} & \ding{51} & \ding{51} & \ding{51} \\
\bottomrule
\end{tabular}
\end{sc}
\end{small}
\end{center}
\end{table}

\paragraph{Image and Layout Generation.} 
\textbf{Image generation} models~\cite{ho2020denoising, nichol2021improved, rombach2022high, song2020denoising, zhang2023adding, zhang2025creatidesign, gao2025high} excel in fidelity but produce non-editable raster outputs. Conversely, \textbf{layout generation} methods~\cite{li2019layoutgan, jyothi2019layoutvae, gupta2021layouttransformer, arroyo2021variational, feng2023layoutgpt, lin2023layoutprompter, seol2024posterllama, yang2024posterllava, cheng2025graphic} focus on abstract spatial arrangements, leaving visual realization to manual assembly.
To bridge the gap between layout planning and visual creation, recent \textbf{combinations of image and layout generation}~\cite{chen2025posta, qu2025igd, zhang2025creatiposter, wei2025postercopilot,  wang2025banneragency, liu2026posterverse} aim to produce editable designs by integrating asset generation with layout compositing.
However, as summarized in Table~\ref{tab:comparison}, existing methods fail to simultaneously address the critical requirements of professional design.
While specific approaches like IGD~\cite{qu2025igd} achieve a \textit{Flexible Schema} by breaking free from fixed background-foreground templates via sequential layer generation, and others such as PosterCopilot~\cite{wei2025postercopilot} and BannerAgency~\cite{wang2025banneragency} focus on reflection mechanisms to ensure \textit{Visual Coherence}, they often overlook the complexity of element styling.
Conversely, although PosterVerse~\cite{liu2026posterverse} successfully incorporates \textit{Expressive Representation}, it lacks the structural flexibility and global harmony found in other specialized methods.
Consequently, no prior work successfully integrates all three capabilities within a single system.
Our method is the first to bridge these gaps within a unified framework. A comparison of our programmatic synthesis against traditional Text-to-Image and layout generation approaches is further illustrated in Figure~\ref{fig:3_main_approaches}.

\paragraph{Multi-Layer Image Generation and Decomposition.}
To address the non-editable nature of flat images, some recent works~\cite{chen2025rethinking, suzuki2025layerd, liu2025omnipsd, yin2025qwen, zhao2025utdesign} focus on generating or decomposing designs into separated RGBA layers. However, they remain raster-bound, lacking the parametric control (e.g., editable fonts) inherent to code.

\paragraph{Code Generation for Design.} 
While code-driven methods have advanced UI reconstruction~\cite{wu2025mllm, chen2025designcoder, laurenccon2024unlocking, chen2025psd2code} and vector graphics~\cite{rodriguez2023starvector, xing2024svgdreamer, xing2025svgdreamer++}, their extension into creative graphic design remains superficial. Existing works~\cite{tang2023layoutnuwa, seol2024posterllama, kikuchi2025multimodal, hsu2025postero} primarily utilize code as a structural carrier. In contrast, \textbf{DesignAsCode} leverage full HTML/CSS for fine-grained, renderable, and stylistically rich design synthesis.

\begin{figure*}[t]
  \centering
  \includegraphics[
    width=\textwidth,
    trim=0 250 0 0,
    clip
  ]{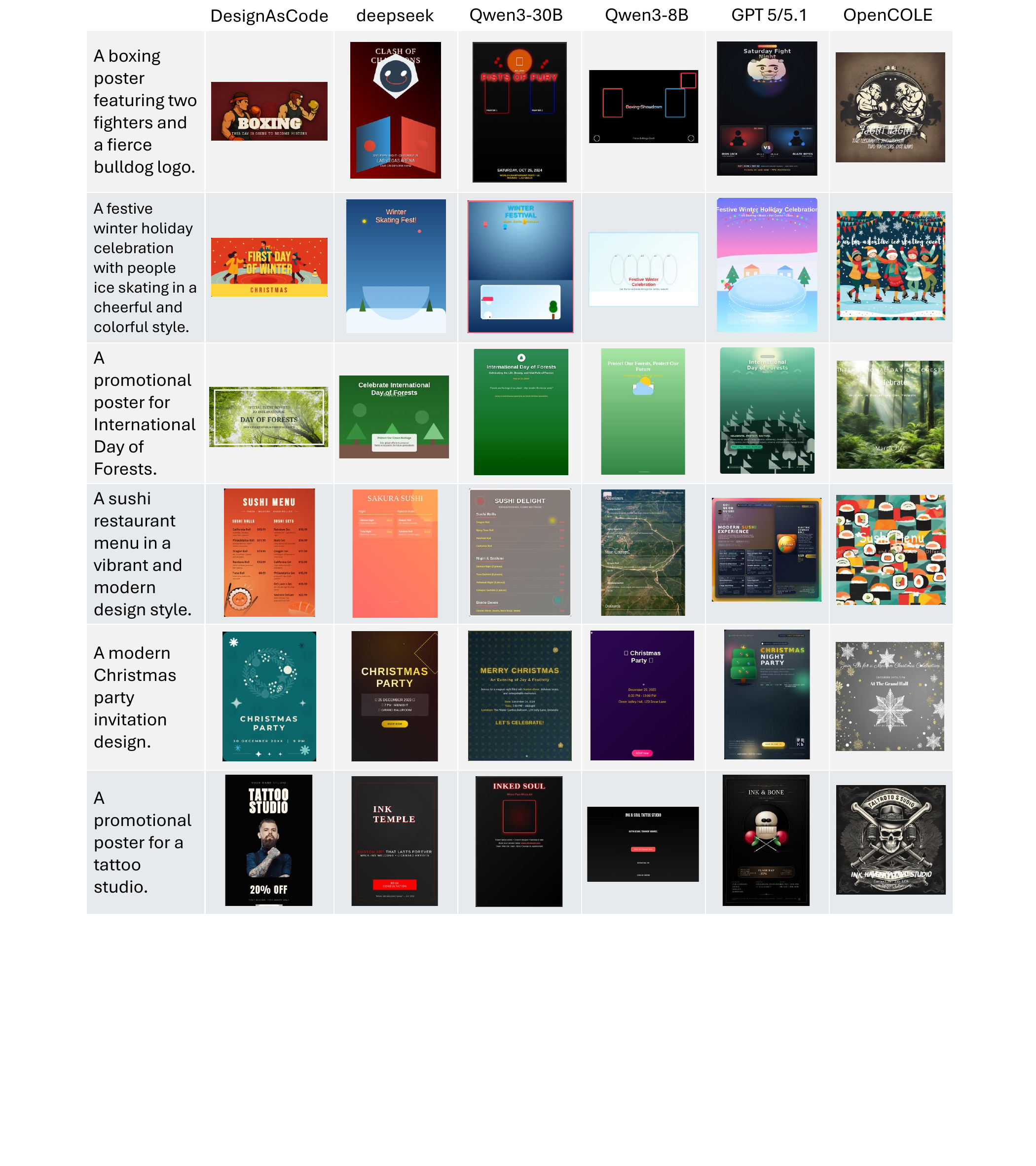}
  \caption{\textbf{Qualitative experimental results.} The leftmost column displays the input text prompts. The top three rows demonstrate the results on the Crello test set, while the bottom three rows show the results on the Broad test set.}
  \Description{This image displays a grid comparing the graphic design layouts generated by six different methods across six diverse text prompts, illustrating that the proposed DesignAsCode approach consistently produces more visually cohesive and structurally complete designs than the baseline models.}
  \label{fig:quality}
\end{figure*}

\section{Methodology}

Given a single-sentence user instruction, our pipeline synthesizes a high-quality editable graphic design.
We adopt HTML/CSS as the underlying representation, as introduced in \S\ref{sec:representation}.
To achieve this, we propose a \textit{Plan-Implement-Reflect} framework, which corresponds to the three subsequent sections: the Semantic Planner (\S\ref{sec:planner}), HTML Implementation (\S\ref{sec:implementation}), and Visual-Aware Reflection (\S\ref{sec:reflection}).

\subsection{Design Generation as HTML Code Generation}
\label{sec:representation}

We formalize graphic design generation as a structured code synthesis problem where a user instruction $I$ is mapped to a design $D$ via $I \to D$. Here, the design is represented as executable HTML/CSS code defined by $D = \mathcal{H}(\{ \textit{img}_i \}_{i=1}^{M}, \{ \textit{text}_j \}_{j=1}^{N})$, with $\mathcal{H}(\cdot)$ denoting an HTML document that composes a set of visual assets $\{\textit{img}_i\}$ and textual elements $\{\textit{text}_j\}$ into a coherent layout.
Here, each element is associated with explicit structural hierarchy, spatial positioning, and style attributes defined via standard HTML tags and CSS rules.
Importantly, we do \emph{not} impose any predefined templates, restricted tag sets, or bounded CSS styles on the generated code.
The model is free to synthesize arbitrary HTML structures and CSS properties, including nested layouts, flexible box or grid systems, layered positioning, and diverse typographic or visual styles.
This unconstrained representation allows the system to fully exploit the expressive power of web standards, avoiding the rigidity and compositional bias introduced by fixed layout schemas or hand-crafted design grammars.

\subsection{Semantic Planner}
\label{sec:planner}

The Semantic Planner module accepts a single-sentence natural language prompt as input and sequentially synthesizes four distinct components. First, a \textbf{layout reasoning sequence} articulates the chain-of-thought (CoT)~\cite{wei2022chain} process of design planning, linguistic layout descriptions, and general element content. Second, \textbf{element groupings} categorize components based on spatial and semantic attributes to guide subsequent generation steps. Third, \textbf{image layer prompts} are generated specifically for each image layer. Finally, \textbf{textual specifications} define details such as content, font size, typeface, and color.

To equip our Semantic Planner with professional design expertise, we employ a knowledge distillation strategy to extract high-level design reasoning and content planning logic from real-world data.
Constructing the training corpus based on the Crello dataset~\cite{yamaguchi2021canvasvae}, which offers rich design meta-information including individual layers, font attributes, and layout specifications for each image, involves reverse-engineering the design process through advanced teacher models.
First, we utilize GPT-4o~\cite{hurst2024gpt} to synthesize simulated single-sentence user instructions from the final design renderings, while concurrently extracting \textit{layout reasoning sequences}, \textit{element groupings}, and \textit{textual specifications} from the combined analysis of the design metadata and visual layouts.
For the specific task of describing visual assets, we employ GPT-o3 to distill precise \textit{image layer prompts} from individual image layers.
Finally, we fine-tune a lightweight LLM (Qwen3-8B~\cite{yang2025qwen3}) on this composite dataset.
This Supervised Fine-Tuning (SFT) process is designed to elicit the model's capability to translate abstract user instructions into structured intermediate plans comprising spatial logic and asset descriptions prior to any implementation in code.

\subsection{HTML Implementation}
\label{sec:implementation}

This module instantiates the abstract plan into an executable design through a two-step process, effectively utilizing the structured outputs from the Semantic Planner.

First, the \textit{image layer prompts} are employed to acquire the necessary visual assets.
This acquisition operates via a flexible channel: the prompts either serve as queries to retrieve semantically matching assets from a repository we constructed or directly act as generative instructions for a Text-to-Image model (e.g., GPT-image-1) to synthesize specific layers.

Second, the obtained visual assets are aggregated with the Planner's remaining structural constraints, including the \textit{element groupings} and \textit{textual specifications}.
These combined inputs are fed into a state-of-the-art Multimodal LLM (e.g., GPT-5), which serves as the layout Composer.
Note that to optimize token efficiency, visual assets are provided as URL references paired with their semantic descriptions rather than raw pixel data.
Leveraging its code-generation capabilities, the model synthesizes the initial HTML/CSS document, ensuring that the retrieved/generated images are correctly linked and the text elements are rendered with appropriate styles according to the planned hierarchy.

\subsection{Visual-Aware Reflection}
\label{sec:reflection}

Although the initially generated HTML establishes a functional structure, it often harbors the \textit{visual dissonance} highlighted in our Introduction.
Crucially, identifying these aesthetic defects poses a fundamental challenge for text-only code generators.
Many visual incoherencies remain invisible in the raw HTML/CSS representation and only manifest when the code is rendered and composited with actual image assets.
To bridge this perception gap, we propose HTML Optimization via Visual-Aware Reflection.
By integrating a visual perception module that evaluates and refines the rendered design, this framework enables the system to detect and rectify subtle aesthetic flaws that are indiscernible from the code alone.

Specifically, following the generation of the initial HTML, the system initiates a visual-aware optimization process. 
During this stage: (1) the current HTML is rendered into a rasterized image \textit{via} an automated browser engine (Playwright); (2) an image editing model (instantiated with GPT-image-1 in our implementation) optimizes this rendering to enhance its visual quality; and (3) the Refiner (utilizes the same model as the Composer) refines the underlying HTML code by conditioning on the comparison between the optimized image, the original rendering, and the current HTML structure.

This Visual-Aware Reflection mechanism strategically exploits the robust aesthetic capabilities of image generation models to guide structural optimization, thereby circumventing the inherent non-editability of raster outputs. Through this mechanism, the system effectively enhances layout alignment and stylistic consistency, resulting in a refined and polished HTML design.

\begin{table*}[t]
\centering
\caption{\textbf{Quantitative comparison on Standard (Crello) and Broad-Spectrum (Broad) Benchmarks.} Our method outperforms baselines in structural validity and visual fidelity across both datasets.}
\label{tab:combined_results}
\setlength{\tabcolsep}{4pt}

\begin{tabular*}{\textwidth}{
  @{\extracolsep{\fill}}
  llcccccccc
  @{}
}
\toprule
\multirow{2}{*}{\textbf{Dataset}} & \multirow{2}{*}{\textbf{Method}} & \multicolumn{4}{c}{\textbf{Objective Metrics}} & \multicolumn{4}{c}{\textbf{Subjective Metrics}} \\ \cmidrule(lr){3-6} \cmidrule(lr){7-10} 
 & & \textbf{Val} $\uparrow$ & \textbf{Ali} $\downarrow$ & \textbf{Rea} $\downarrow$ & \textbf{Clip} $\uparrow$ & \textbf{Text} $\uparrow$ & \textbf{Image} $\uparrow$ & \textbf{Layout} $\uparrow$ & \textbf{Color} $\uparrow$ \\ \midrule
 
\multirow{6}{*}{\textbf{Crello}} 
 & \textbf{DesignAsCode (Ours)} & \textbf{0.9521} & \underline{0.0008} & \textbf{0.0849} & \textbf{0.6287} & \textbf{74.06} & \textbf{65.86} & \textbf{63.67} & \textbf{68.42} \\
 & DeepSeek-R1 & \underline{0.9224} & 0.0040 & \underline{0.0970} & 0.5478 & 71.08 & 37.10 & 57.47 & 61.03 \\
 & Qwen3-30B & 0.8930 & 0.0057 & 0.0989 & 0.5386 & 68.22 & 35.12 & 55.60 & 57.98 \\
 & Qwen3-8B & 0.9131 & 0.0031 & 0.1152 & 0.5196 & 57.62 & 34.25 & 41.14 & 49.89 \\
 & GPT-5 & 0.6931 & \textbf{0.0005} & 0.1466 & 0.5669 & \underline{73.51} & 41.26 & \underline{57.19} & \underline{65.48} \\
 & OpenCOLE & -- & -- & -- & \underline{0.5864} & 48.57 & \underline{65.09} & 56.30 & 53.63 \\ 
\midrule

\multirow{6}{*}{\textbf{Broad}} 
 & \textbf{DesignAsCode (Ours)} & \textbf{0.9905} & \underline{0.0003} & \textbf{0.0911} & \textbf{0.6732} & \underline{68.63} & \textbf{54.06} & \textbf{55.46} & \textbf{67.63} \\
 & DeepSeek-R1 & 0.9881 & 0.0026 & 0.1031 & 0.6205 & 60.08 & 34.08 & 42.88 & \underline{58.92} \\
 & Qwen3-30B & 0.9714 & 0.0012 & \underline{0.0983} & 0.6213 & 60.84 & 29.18 & 41.04 & 57.88 \\
 & Qwen3-8B & \underline{0.9892} & 0.0016 & 0.1338 & 0.5911 & 46.32 & 30.72 & 30.96 & 50.32 \\
 & GPT-5.1 & 0.9458 & \textbf{0.0001} & 0.1219 & \underline{0.6226} & \textbf{75.96} & 44.56 & \underline{50.76} & 57.80 \\
 & OpenCOLE & -- & -- & -- & 0.6108 & 46.44 & \underline{53.68} & 48.64 & 56.00 \\ 
\bottomrule
\end{tabular*}

\end{table*}

\begin{table}[t]
    \centering
    \caption{\textbf{User Study Results (Average Rank).} Comparison of Structure Rationality and Visual Aesthetics. The Average Rank is reported ($\downarrow$ lower is better).}
    \label{tab:user_study}
    \resizebox{\linewidth}{!}{ 
    \begin{tabular}{lcc}
        \toprule
        \textbf{Method} & \textbf{Structure Rationality} & \textbf{Visual Aesthetics} \\
         & \textbf{Avg. Rank} ($\downarrow$) & \textbf{Avg. Rank} ($\downarrow$) \\
        \midrule
        \textbf{DesignAsCode} & \textbf{1.92} & \textbf{2.03} \\
        DeepSeek-R1 & 3.38 & 3.70 \\
        Qwen3-30B   & 3.57 & 4.66 \\
        Qwen3-8B    & 4.35 & 4.90 \\
        GPT-5       & 3.87 & 3.28 \\
        OpenCOLE    & 3.91 & 2.44 \\
        \bottomrule
    \end{tabular}
    }
    \vspace{-5pt}
\end{table}

\begin{table*}[t]
\centering
\caption{\textbf{Ablation Studies on the Crello Dataset.} \textbf{Full} denotes our complete DesignAsCode framework. \textbf{w/o HTML Representation} removes the HTML representation and use JSON as an alternative. \textbf{w/o Semantic Planner} replaces our specialized planner with a general-purpose LLM (GPT-5). \textbf{w/o Visual Reflection} removes the Visual-Aware Reflection mechanism.}
\label{tab:ablation}

\begin{tabular*}{\textwidth}{
  @{\extracolsep{\fill}}
  llcccccccc
  @{}
}
\toprule
\multirow{2}{*}{\textbf{Method}} & \multicolumn{4}{c}{\textbf{Objective Metrics}} & \multicolumn{4}{c}{\textbf{Subjective Metrics}} \\ \cmidrule(lr){2-5} \cmidrule(lr){6-9} 
 & \textbf{Val} $\uparrow$ & \textbf{Ali} $\downarrow$ & \textbf{Rea} $\downarrow$ & \textbf{Clip} $\uparrow$ & \textbf{Text} $\uparrow$ & \textbf{Image} $\uparrow$ & \textbf{Layout} $\uparrow$ & \textbf{Color} $\uparrow$ \\ \midrule
\textbf{DesignAsCode (Full)} & 0.9521 & 0.0008 & 0.0849 & 0.6287 & 74.06 & 65.86 & 63.67 & 68.42 \\
w/o HTML Representation & 0.9559 & 0.0030 & 0.0878 & 0.6210 & 55.35 & 64.24 & 58.07 & 53.91 \\
w/o Semantic Planner & 0.9389 & 0.0004 & 0.0950 & 0.5896 & 67.62 & 61.14 & 52.78 & 64.25 \\
w/o Visual Reflection & 0.9357 & 0.0025 & 0.0942 & 0.6299 & 64.21 & 62.64 & 58.97 & 60.99 \\ \bottomrule
\end{tabular*}

\end{table*}

\section{Experiments}
\subsection{Experimental Setup}

\paragraph{Datasets and Benchmarks.}
We strictly distinguish between training and testing inputs. The \textbf{Training Set} follows the pipeline in Sec.~\ref{sec:planner}, supervising the model with paired \textit{text prompts} and \textit{layout reasoning sequence, element groupings, image layer prompts, textual specifications}. In contrast, the \textbf{Test Set} inputs are only \textit{text prompts}.
We evaluate on two distinct sets: 1) The \textbf{Standard Benchmark (Crello)}~\cite{yamaguchi2021canvasvae} (sampled 546 cases), serving as the primary basis for reproducible and fair comparisons; and 2) The \textbf{Broad-Spectrum Benchmark (Broad)}, comprising 500 high-quality samples covering a wider variety of categories (e.g., menus, business timelines). We include Broad additionally to demonstrate our model's generalizability to highly diverse, real-world scenarios beyond standard datasets. While the original designs for Broad are proprietary, we release the constructed text prompt benchmarks to provide the community with an extra challenging testbed, ensuring this supplementary evaluation does not hinder the overall reproducibility established on Crello.

\paragraph{Baselines.}
We benchmark against state-of-the-art HTML/CSS generators, including DeepSeek-R1~\cite{guo2025deepseek} and Qwen3-8B/30B~\cite{yang2025qwen3}, utilizing \textbf{GPT-5} and \textbf{GPT-5.1} for the Standard and Broad-Spectrum benchmarks, respectively. We also compare with OpenCOLE~\cite{inoue2024opencole}, excluding other works in Table~\ref{tab:comparison} due to non-open-sourced code.

\paragraph{Evaluation Metrics.}
We adopt a hybrid evaluation protocol comprising objective rule-based metrics, subjective perceptional scores, and human evaluation.
\textbf{1) Objective Metrics.}
To quantitatively assess the layout and visual quality, we employ four metrics:
\textbf{Validity (Val $\uparrow$)} checks the integrity of generated elements;
\textbf{Alignment (Ali $\downarrow$)} measures the layout regularity by calculating spatial deviations between adjacent elements;
\textbf{Readability (Rea $\downarrow$)} assesses text legibility by detecting background clutter (gradients) within text regions;
and \textbf{CLIP Score (Clip $\uparrow$)} evaluates the semantic and stylistic fidelity between the generated output and the ground truth.
\textbf{2) Subjective Metrics.}
To align with human aesthetic perception, we utilize an MLLM judge (GPT-4.1-mini) following~\cite{ge2025autopresent}. The model scores the designs on a scale of 0--100 across four dimensions: \textbf{Text}, \textbf{Image}, \textbf{Layout}, and \textbf{Color}.
\textbf{3) Human Evaluation.} 
To further assess our generated designs through human perception, we focus on two core dimensions: \textit{Structure Rationality} and \textit{Visual Aesthetics}. The detailed experimental protocol and quantitative results of this user study are presented in \S~\ref{sec:user_study}.

\subsection{Main Results and Analysis}
\label{sec:main_results}

We evaluate DesignAsCode against baselines on Standard (Crello) and Broad-Spectrum (Broad) benchmarks via quantitative metrics (Table~\ref{tab:combined_results}) and qualitative samples (Figure~\ref{fig:quality}).

\paragraph{Performance on Standard Benchmark.} 
Quantitative results (Table~\ref{tab:combined_results}, Top) show our method establishes a clear lead across all metrics except Alignment. While ranking second in Alignment, our score reaches a negligible magnitude comparable to the top-performing GPT-5. Qualitatively (Figure~\ref{fig:quality}, top 3 rows), our alignment is indistinguishable from GPT-5, yet we significantly outperform all baselines in aesthetics and visual fidelity.

\paragraph{Performance on Broad-Spectrum Benchmark.} 
Quantitative results (Table~\ref{tab:combined_results}, Bottom) show our method leads in six metrics: Validity, Readability, Clip, Image, Layout, and Color. Although ranking second to GPT-5.1 in subjective Text scores, we significantly outperform it in all other subjective metrics. Qualitative analysis (Figure~\ref{fig:quality}, bottom) attributes GPT-5.1's high Text score to a web-UI bias---relying on rigid containers and reduced complexity (e.g., excessive whitespace) for alignment and text clarity. This approach diverges from the organic text-image integration fundamental to professional posters, resulting in poor Layout and Color performance. Conversely, our \textbf{Semantic Planner} resolves this web-UI bias by design, achieving the best overall quality with only a marginal trade-off in Text scores.

\subsection{User Study}
\label{sec:user_study}
To perceptually evaluate the aforementioned \textit{Structure Rationality} and \textit{Visual Aesthetics}, we conducted a targeted user study. We curated 10 distinct design prompts and generated the corresponding outputs using all six methods. We recruited 20 participants with graphic design experience to rank the anonymized and randomized outputs from 1 (best) to 6 (worst).  

As shown in Table~\ref{tab:user_study}, DesignAsCode consistently achieved the best (lowest) average rank across both dimensions. To validate the statistical significance, we performed a Wilcoxon signed-rank test comparing our method against the second-best performing baseline in each respective metric. The analysis confirmed that the performance improvements of DesignAsCode are statistically significant for both Structure Rationality ($p < 0.001$, vs. DeepSeek-R1) and Visual Aesthetics ($p < 0.05$, vs. OpenCOLE). These compelling perceptual results further demonstrate our method's superiority in maintaining structural integrity without compromising visual quality.

\subsection{Ablation Studies}
\label{sec:ablation}

Table~\ref{tab:ablation} summarizes the ablation studies on the Crello dataset, validating the effectiveness of each module.

\paragraph{Impact of HTML Representation.}
Replacing HTML/CSS with JSON degrades structural precision (Ali: $0.0008 \to 0.0030$) and severely impacts subjective metrics. This confirms that primitive representations lack the expressiveness required for advanced styling (e.g., \textit{backdrop-filter}), failing to resolve complex visual dissonances that native CSS handles effectively.

\paragraph{Impact of Semantic Planner.}
While replacing our planner with a general LLM (GPT-5) yields the lowest alignment error, our method maintains highly comparable precision. However, GPT-5 incurs a steep aesthetic cost (Layout: $63.67 \to 52.78$; Clip: $0.6287 \to 0.5896$). Trained on professional datasets, our Semantic Planner generates superior design plans, transcending the constraints of general LLMs that overly prioritize rigid alignment at the expense of professional planning capabilities.

\paragraph{Impact of Visual-Aware Reflection.}
Removing the reflection mechanism leads to a significant degradation across most metrics. This demonstrates that vanilla code generation is inherently an ``open-loop'' system, blind to rendering artifacts. Our reflection module addresses this by integrating visual feedback into the synthesis process, acting as an ``Art Director'' that perceives and resolves pixel-level conflicts to ensure visual harmony.

\begin{figure}[t]
    \centering
    \includegraphics[
    width=\columnwidth,
    trim=100 30 100 0,
    clip
  ]{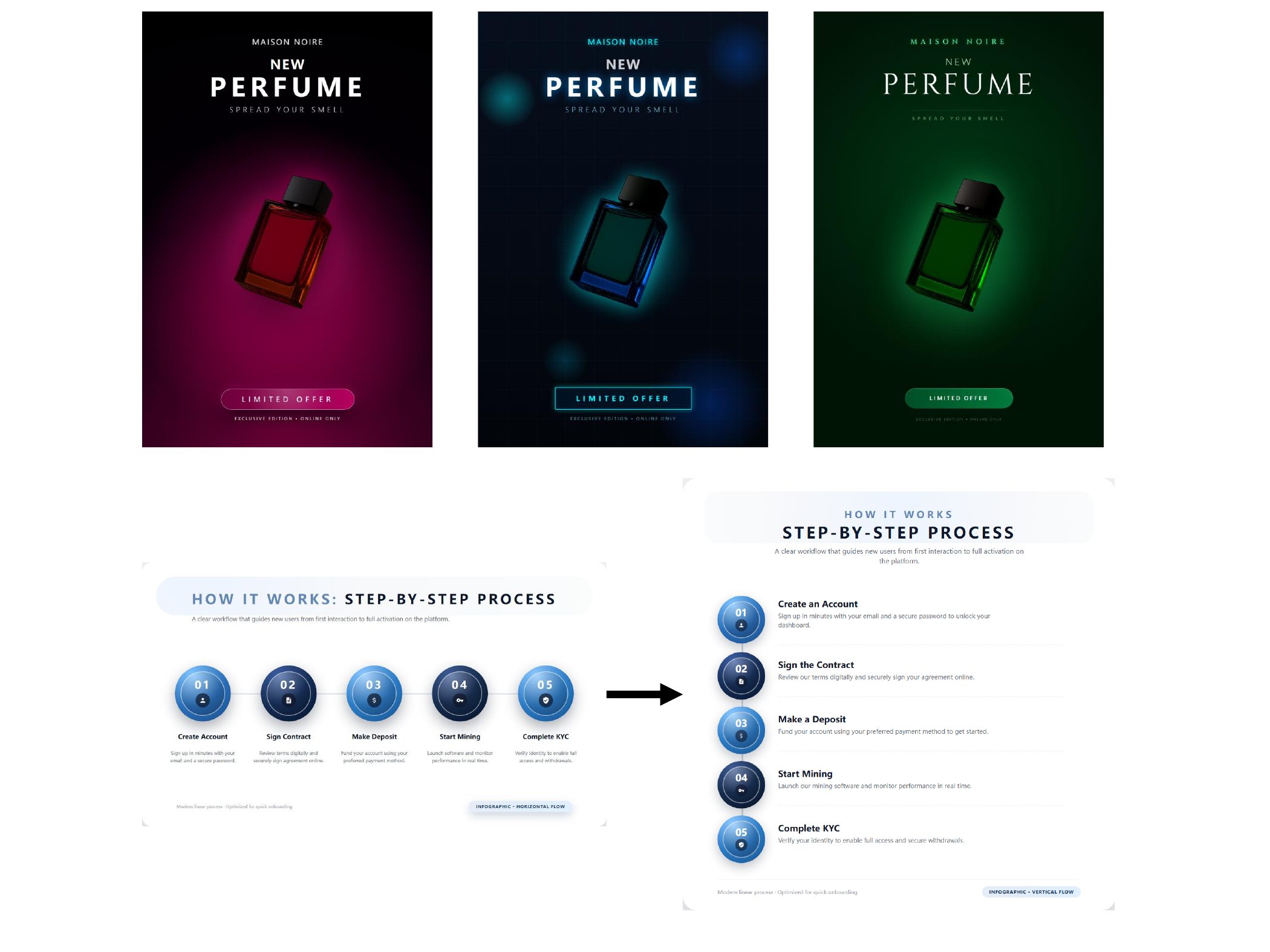}
    \caption{\textbf{Structural Editing and Adaptation.} Code generation enables (Top) global theme switching via CSS variables and (Bottom) content-aware layout retargeting, where elements automatically reflow to fit new aspect ratios.}
    \Description{This image demonstrates the flexibility of code-based design generation, showing how a single design can easily switch between red, blue, and green color themes at the top, and how a horizontal process layout automatically reorganizes into a vertical stack when the aspect ratio changes at the bottom.}
    \label{fig:editing_reframe}
\end{figure}

\begin{figure}[t]
    \centering
    \includegraphics[
    width=\columnwidth,
    trim=30 160 30 0,
    clip
  ]{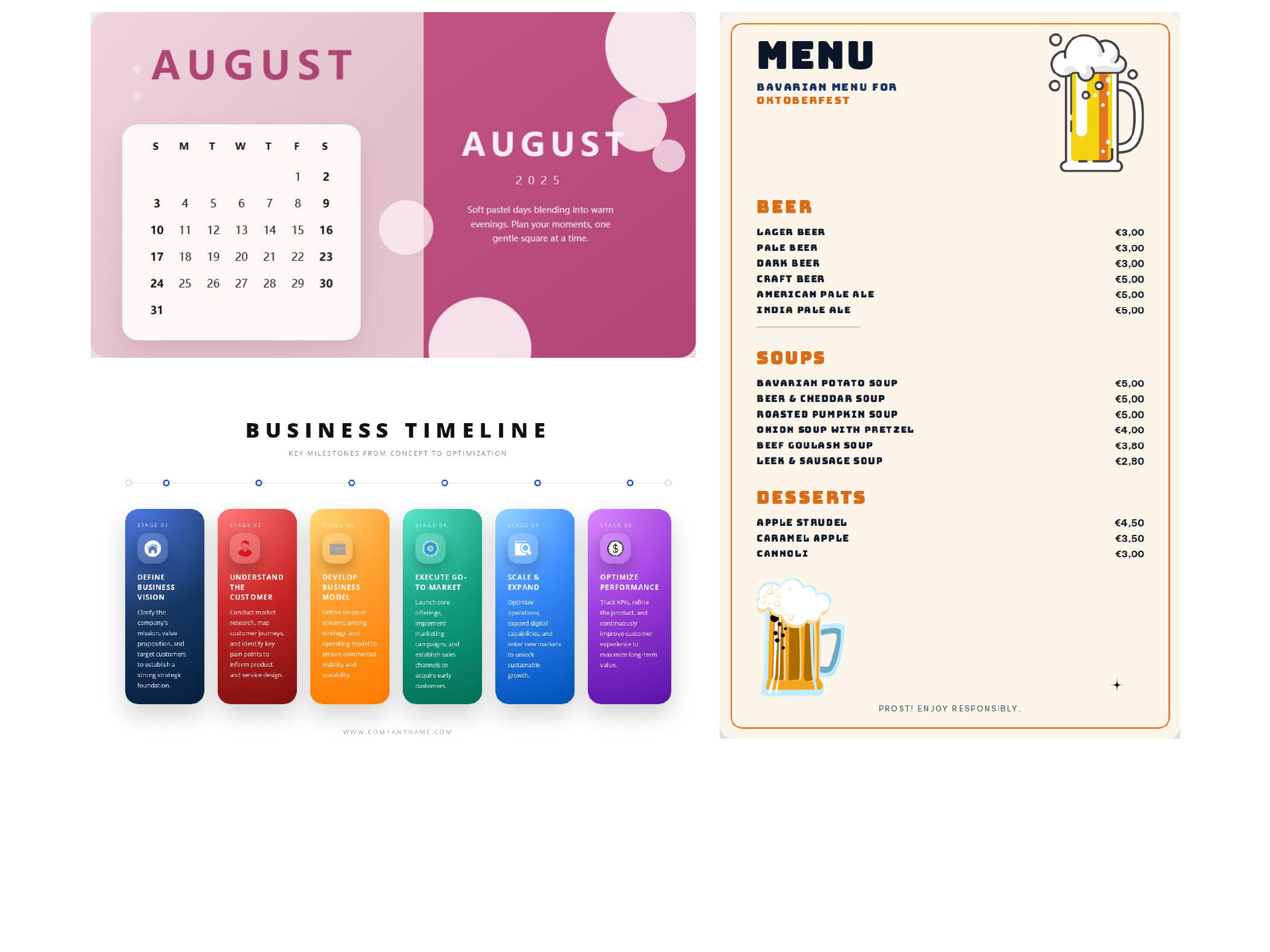}
    \caption{\textbf{Handling High Information Density.} Top-Left: A \textit{Calendar} demonstrating precise grid alignment. Bottom-Left: A \textit{Business Timeline} visualizing a multi-step process. Right: A \textit{Menu} maintaining rigorous price-list alignment.}
    \Description{This image presents three examples of information---dense layouts—a grid-aligned calendar, a multi-step business timeline, and a column-aligned restaurant menu---demonstrating the system's capability to generate complex, text-heavy designs with precise spatial arrangements.}
    \label{fig:complex_layouts}
    
\end{figure}

\begin{figure}[t]
  \centering
  \includegraphics[
    width=\columnwidth,
    trim=30 160 30 0,
    clip
  ]{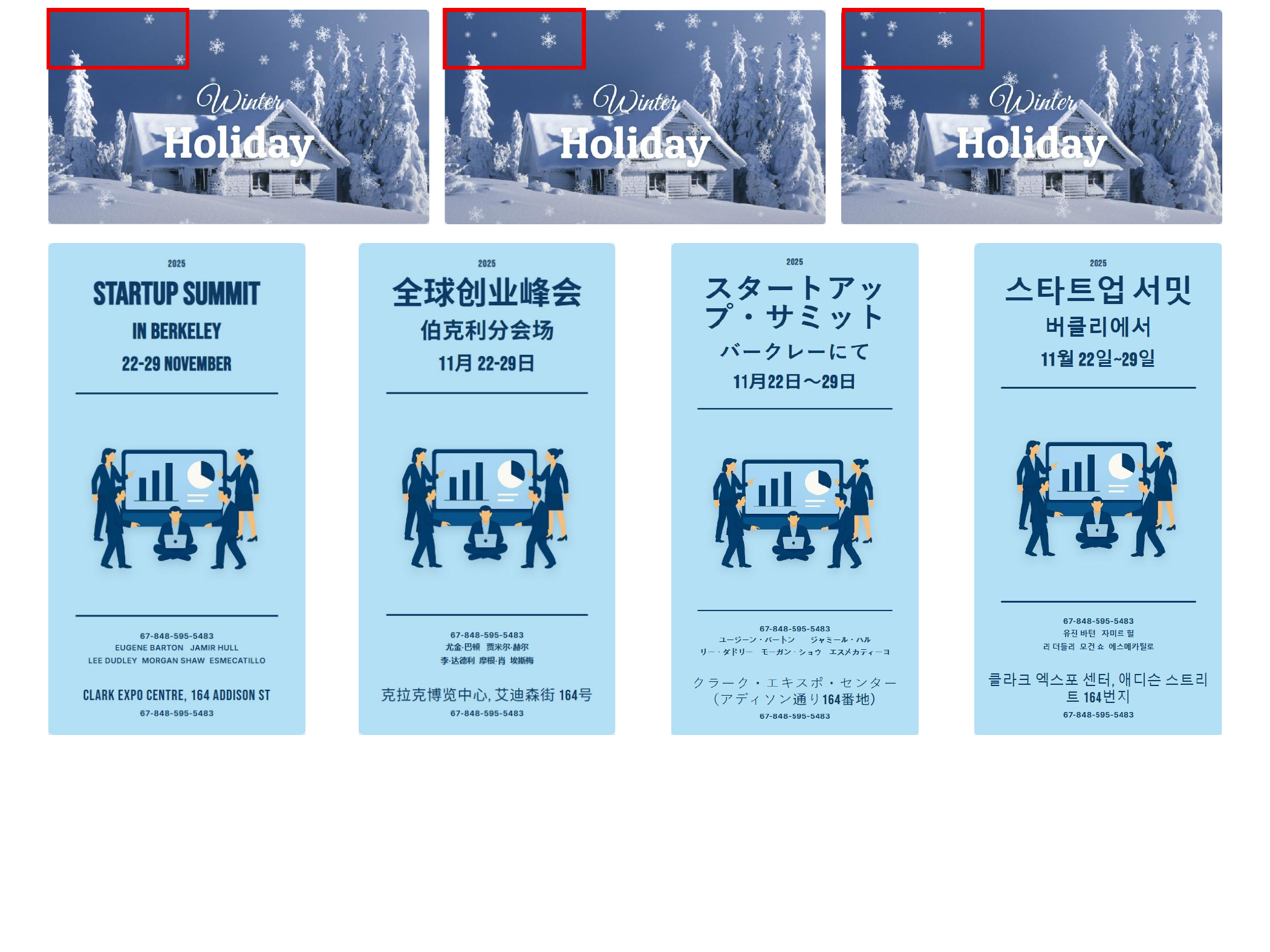}
  \caption{\textbf{Code-Native Extensions.} Top: A filmstrip visualization of an animated poster generated via CSS keyframes. Significant visual changes are visible within the red box. Bottom: The model's ability to render perfect text in multiple languages (English, Chinese, Japanese, Korean).}
  \Description{This image illustrates code-native design extensions by showing a sequence of frames from an animated winter poster with moving snowflakes highlighted in red boxes at the top , alongside four identical summit poster layouts perfectly rendered in English, Chinese, Japanese, and Korean at the bottom.}
  \label{fig:extensions}
\end{figure}

\section{Applications}

Leveraging the HTML/CSS representation and the Plan-Implement-Reflect pipeline, DesignAsCode enables diverse downstream applications.

\paragraph{Attribute-Level Editability.} Disentangled code allows instantaneous modification of global attributes (e.g., themes) or local properties via CSS variables without artifacts (Figure~\ref{fig:editing_reframe}, Top).

\paragraph{Layout Retargeting.} Leveraging responsive CSS achieves intrinsic retargeting. Resizing the container triggers automatic layout reflow---reorganizing elements (e.g., vertical stacking) to fit new aspect ratios while preserving design integrity (Figure~\ref{fig:editing_reframe}, Bottom), solving the distortion issues common in pixel scaling.

\paragraph{Complex Information Layouts.} Existing hybrid methods like PosterCopilot~\cite{wei2025postercopilot} and PosterVerse~\cite{liu2026posterverse} often fail to balance visual fidelity with high information density required for text-dense documents. By leveraging our Plan-Implement-Reflect pipeline, DesignAsCode synthesizes high-density layouts with rigorous typographic hierarchy. Figure~\ref{fig:complex_layouts} demonstrates diverse scenarios, from grid-based calendars to detailed menus, ensuring strict spatial alignment while accommodating large volumes of textual information.

\paragraph{Animated Posters.} We can extend static designs into dynamic media by prompting the model to inject CSS keyframes (e.g., falling snow), adding visual appeal without external video models (Figure~\ref{fig:extensions}, Top).

\paragraph{Multilingual Generation.} Our approach decouples content from rendering. By updating text strings and font-families in the code, we achieve perfect multilingual output while maintaining the original design structure, avoiding the text artifacts common in diffusion models (Figure~\ref{fig:extensions}, Bottom).

\paragraph{Dynamic Data Binding and Batch Personalization.}
By treating the generated HTML/CSS as a programmatic template, DesignAsCode enables seamless integration with backend databases. Variables such as user names, real-time pricing, or localized imagery can be dynamically injected into the code layout. This facilitates large-scale, automated batch personalization for e-commerce campaigns without requiring repeated model inference or manual redesign.

\paragraph{Accessibility (A11y) and SEO Compatibility.}
A fundamental limitation of rasterized posters is their opacity to screen readers and search engines. Because our designs are represented as semantic HTML, the textual content remains fully machine-readable. This ensures that the generated graphics are inherently accessible to visually impaired users via screen readers and allows search engine crawlers to index the embedded textual information—a crucial advantage for real-world digital marketing.

\section{Conclusion}

\textbf{DesignAsCode} bridges structural editability and visual fidelity by reimagining graphic design as HTML/CSS synthesis. Our \textit{Plan-Implement-Reflect} pipeline integrates semantic planning with visual-aware reflection to ensure professional-grade outputs. By shifting from pixel mimicry to executable semantic logic, our method outperforms baselines and unlocks a vast range of applications, enabling structural manipulation and opening new directions for controllable, high-fidelity design generation.


\balance
\bibliographystyle{ACM-Reference-Format}
\bibliography{main}

\clearpage
\onecolumn
\appendix

\section{Implementation Details}
\subsection{Semantic Planner Training Details}
\label{apd:training}

We trained two distinct planners to address standard tasks from the public Crello~\cite{yamaguchi2021canvasvae} dataset and complex tasks from a private dataset, respectively. For the planner handling standard tasks, we utilized a subset of approximately 10,000 samples, representing half of the Crello training set. For the planner designed for complex tasks, we combined the aforementioned 10,000 Crello samples with an additional 7,500 samples selected from our private dataset. Detailed training configurations are provided in Table~\ref{tab:training_details}.

\begin{table}[ht]
\caption{Hyperparameters and implementation details for fine-tuning semantic planner.}
\label{tab:training_details}
\centering
\begin{tabular}{lll}
\toprule
\textbf{Category} & \textbf{Hyperparameter} & \textbf{Value} \\
\midrule
\multirow{3}{*}{\textbf{Model}} 
 & Base Model & Qwen/Qwen3-8B~\cite{yang2025qwen3} \\
 & Architecture Type & Causal Decoder-only \\
 & Training Method & Full Fine-tuning (FFT) \\
\midrule
\multirow{7}{*}{\textbf{Optimization}} 
 & Optimizer & AdamW~\cite{loshchilov2017decoupled} \\
 & Learning Rate & $5 \times 10^{-5}$ \\
 & Learning Rate Scheduler & Linear Decay (Default) \\
 & Batch Size (Per Device) & 1 \\
 & Gradient Accumulation & 2 Steps \\
 & \textbf{Global Batch Size} & \textbf{16} ($1 \times 8 \text{ GPUs} \times 2$) \\
 & Epochs & 2 \\
\midrule
\multirow{3}{*}{\textbf{Data \& Loss}} 
 & Max Sequence Length & 6144 tokens \\
 & Format & ChatML (System + User + Assistant) \\
 & Loss Objective & Completion Only (Masked User Prompt and System Prompt) \\
\midrule
\multirow{3}{*}{\textbf{Infrastructure}} 
 & GPUs & 8 $\times$ NVIDIA A100 (40GB) \\
 & Distributed Strategy & DeepSpeed (ZeRO Stage 3)~\cite{rasley2020deepspeed} \\
 & Precision & BF16 \\
\bottomrule
\end{tabular}
\end{table}

\subsection{Prompts and Details for Distilling Semantic Plan}
\label{apd:prompt}

\begin{promptbox}{Prompts for Distilling User Input}
\small\ttfamily
Given that this 2D-design image was generated by the user through a model, infer the prompt for generating this image. Note that the user's prompt is just one brief sentence and does not elaborate on what is on the picture, because the user don't exactly know what should on the image. Just return the prompt. No additional content is required. Don't put quotation marks at the beginning or end of the sentence. For example: A menu in the style of a formal event dinner.
\end{promptbox}

\begin{promptbox}{Prompts for Distilling Layout Reasoning Sequence}
\setlength{\parindent}{0pt}
\small\ttfamily
System:

You are an expert in graphic layout design and visual composition. Your task is to analyze the input image layers (including each element’s position, color, size, text, etc.) and the final rendered image, then generate a corresponding `layout\_thought`. The `layout\_thought` is a descriptive text that explains the design thinking process behind the layout—including background setup, visual focal points, text positioning, alignment strategy, and overall visual logic.

\vspace{1em}
User:

Below is the layer information (in JSON format) and the final image. Please analyze the layout logic of the layers and generate a `layout\_thought` that describes the reasoning behind the layer arrangement. The tone should be as if you were planning the layout of this picture rather than explaining its layout. The `layout\_thought` should include:

1. Overall layout strategy (e.g., center alignment, symmetry). Use 'This layout should be like this and that...' more often. Plan with such sentences. 

2. Every layer's element(text or not text), including position, size, opacity, and color, etc. However, there is no need to be particularly specific, because what is generated now is only the chain of thought, and the formal layout will be generated later. You need to plan each floor in sequence from small to large. You need to plan strictly in accordance with the hierarchical sequence, because another model will generate the actual layout plan based on your thought. You must carefully plan each layer, and explain why.

Only output the layout\_thought as plain descriptive text. Do not include explanations, questions, JSON formatting, or Markdown. Just the layout\_thought text.
\end{promptbox}

\begin{promptbox}{Prompts for Distilling Element Grouping}
\setlength{\parindent}{0pt}
\small\ttfamily
System:

You are an expert in graphic layout design and visual composition. Your task is to analyze the input image layers (including each element’s position, color, size, text, etc.) and the final rendered image, then generate a corresponding `grouping`. The `grouping` is a structured plan that defines how related elements (text, shapes, images) should be grouped together. The goal is to identify semantic or visual relationships to guide later layout or generation stages.

\vspace{1em}
User:

Below is the layer information (in JSON format) and the final image.

Please analyze the relationships among the elements and generate the `grouping` result only.

`grouping` identifies semantically or visually related elements that should be treated as a unit in subsequent generation steps. Each group may correspond to a visual block (e.g., a menu item consisting of an image and its text label, a header area, a call-to-action block, etc.).

Each group must be expressed as a JSON object with three fields:

- `group\_id`: a unique identifier string like 'G1', 'G2'...

- `children`: a list of layer\_ids (from the provided JSON) that belong to this group

- `theme`: a short description (2–6 words) summarizing the group's purpose (e.g., 'menu item', 'header block', 'background image').

If there is a stacking relationship among the children in the group, please list them in the order from bottom to top. 

If an element is standalone and not obviously related, it can form its own group.

Strictly follow this format:

<grouping>

[ ... JSON array of groups ... ]

</grouping>

Do not include any other text, explanations, or reasoning outside the <grouping> tags.
\end{promptbox}

\begin{promptbox}{Prompts for Distilling Image Layer Prompts}
\setlength{\parindent}{0pt}
\small\ttfamily
System:

You are an expert prompt reverse-engineer who infers prompts from generated images.

\vspace{1em}
User:

The following image is a layer (element) of a complete 2D design, and it was generated by the model. Below is the structural information of this layer in JSON format, including its position on the origin image, size, etc.

Please provide a prompt for generating this layer (element) as detailed as possible, including information such as content, size, opacity, and shape. But do not include the position information, because the position is determined by the layout of the entire image. Since the image does not contain text, please ignore all text-related information in the JSON. If the 'type' label in JSON is not 2 (ImageElement) or 3 (ColoredBackground), then you need to point out which parts are transparent. Just return the prompt. No additional content is required. Don't put quotation marks at the beginning or end of the sentence.

\end{promptbox}

For the textual specifications, we directly extracted attributes including width, height, opacity, text, font, font\_size, text\_align, angle, capitalize, line\_height, and letter\_spacing from the metadata of each sample in the dataset, without employing large model distillation.

\subsection{Adaptive Retrieval-Generation Mechanism in HTML Implementation}
\label{apd:retrieve}
We construct a layer image repository utilizing visual assets (image layers within complete designs) from the Crello training set, strictly excluding any assets found in the test set, alongside their distilled image prompts. Specifically, we employ a sentence-transformers~\cite{reimers2019sentence} model to map these prompts into a vector embedding space and build a FAISS~\cite{johnson2019billion} index to establish a retrieval mapping to the corresponding images. During the inference phase, once the Planner generates the required visual prompts, the system retrieves the most semantically similar prompts from the index and returns the associated image assets. Subsequently, both the Planner's output and the retrieved images are fed into the Composer (powered by GPT-5) to synthesize the initial HTML design. Following this, a Layer Optimization Agent intervenes. Taking the Planner's specifications, the generated HTML code, and a rendered screenshot as inputs, this agent evaluates whether the image layers satisfy criteria for semantic alignment, stylistic consistency, and composability. If a layer fails to meet these standards, the agent triggers a generative model (GPT-image-1) to regenerate the specific asset. This design leverages the complementarity between retrieval and generation: while image generation offers superior quality and precision, its inference speed is typically slower than retrieval; conversely, retrieval is highly efficient but restricted by the finite nature of the repository, which may not always perfectly fulfill specific semantic requirements. By employing this hybrid, on-demand generation strategy, the mechanism effectively strikes a balance between efficiency and quality. Notably, in extreme scenarios, the system can be configured to operate in either a purely retrieval-based or a purely generative mode. Furthermore, since this process updates the underlying image content without altering the HTML image URLs, the optimized design is obtained immediately without requiring structural reassembly.

\subsection{Algorithm Details of HTML Optimization via Visual-Aware Reflection}
\label{apd:reflection}
Algorithm~\ref{alg:optimization} outlines the detailed procedure.
Currently, the Visual-Aware Reflection functions sequentially as a one-shot process. However, its architecture inherently supports a multi-step iterative paradigm, where the optimized visual representation from the previous stage is seamlessly fed forward as the initial state for the subsequent reflection cycle.

\subsection{Differences between Standard Dataset (Crello) and Broad-Spectrum Dataset (Broad)}
\label{apd:dataset_examples}
\begin{figure}[H]
  \centering
  \includegraphics[
    width=\textwidth,
    trim=0 150 0 0,
    clip
  ]{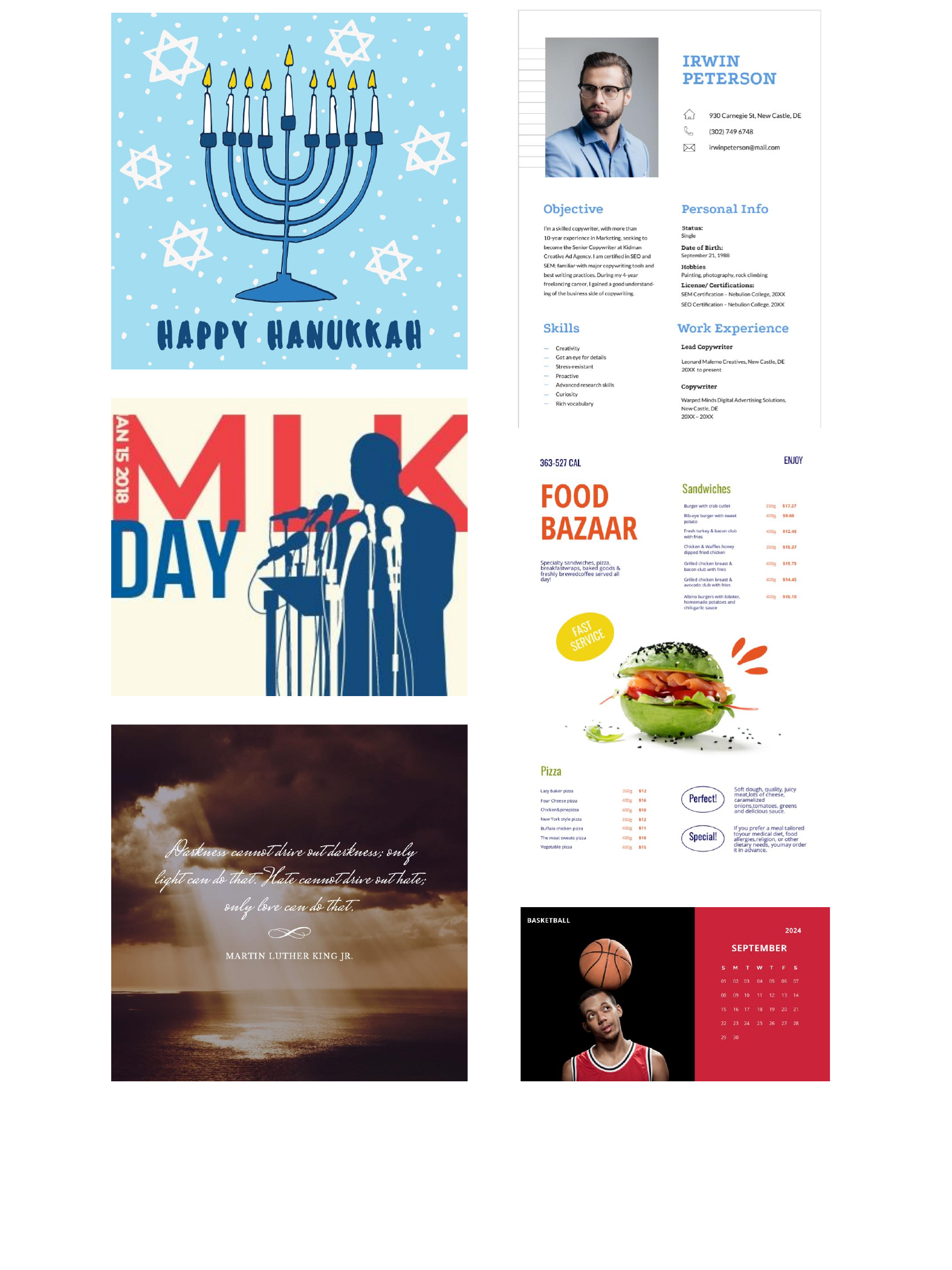}
  \caption{Comparison between the Standard Dataset (left column) and the Broad-Spectrum Dataset (right column).}
  \label{fig:dataset_compare}
\end{figure}

The Broad-Spectrum Dataset encompasses more diverse categories, including menus, mind maps, infographics, invoices, newsletters, timelines, storyboards, resumes, proposals, invitations, brochures, and calendars. Figure~\ref{fig:dataset_compare} qualitatively illustrates the key distinctions between the Standard Dataset and the Broad-Spectrum Dataset.

\begin{algorithm}[tb]
   \caption{Visual-Aware Reflection for HTML Optimization} 
   \label{alg:optimization}
   \begin{algorithmic}[1]
     \STATE {\bfseries Input:} Initial HTML design $\mathcal{H}_{init}$ 
     \STATE {\bfseries Output:} Refined HTML design $\mathcal{H}_{final}$
   
     \STATE \COMMENT{\textbf{Step 1: Render current state}}
     \STATE $I_{render} \leftarrow \text{Render}(\mathcal{H}_{init})$
      
     \STATE \COMMENT{\textbf{Step 2: Visual Optimization (Image Modality)}}
     \STATE \textit{// Image model enhances the visual quality of the rendering}
     \STATE $I_{opt} \leftarrow \text{ImageEditor}(I_{render})$
      
     \STATE \COMMENT{\textbf{Step 3: Structural Refinement (Code Modality)}}
     \STATE \textit{// Refiner updates HTML based on the visual gap}
     \STATE $\mathcal{H}_{final} \leftarrow \text{Refiner}(\mathcal{H}_{init}, I_{render}, I_{opt})$ 
   
     \STATE {\bfseries return} $\mathcal{H}_{final}$
  \end{algorithmic}
\end{algorithm}

\subsection{Evaluation Metrics Details}
\label{apd:metrics}

\subsubsection{Objective Metrics}
Our objective evaluation focuses on structural correctness, geometric precision, and visual consistency.

\noindent\textbf{Validity (Val $\uparrow$).}
This metric quantifies the structural integrity of the generated code by measuring the ratio of valid elements to the total number of generated elements via DOM tree analysis~\cite{hsu2023posterlayout, li2020attribute}.
\begin{equation}
    \text{Val} = \frac{N_{valid}}{N_{total}}
\end{equation}
To filter out imperceptible noise or artifacts, we apply a dynamic area threshold. An element is considered valid only if its bounding box area exceeds $0.1\%$ of the canvas for the Standard Benchmark (as well as the ablation study). For the Broad-Spectrum Benchmark, this threshold is lowered to $0.01\%$ to accommodate fine-grained decorative components.

\noindent\textbf{Alignment (Ali $\downarrow$).}
Alignment measures the layout regularity by quantifying how well elements adhere to a grid system~\cite{hsu2023posterlayout, zhou2022composition, li2020attribute}. We consider the set of valid elements $E = \{e_1, e_2, ..., e_n\}$. For each element $e_i$, we define a set of 6 alignment coordinates $C_i = \{x_{left}, x_{center}, x_{right}, y_{top}, y_{center}, y_{bottom}\}$. 
The alignment score is calculated as the average minimum $L_1$ distance between the coordinates of an element and the coordinates of all other elements, normalized by the canvas diagonal length $D_{canvas}$:
\begin{equation}
    \text{Ali} = \frac{1}{|E| \cdot D_{canvas}} \sum_{e_i \in E} \min_{e_j \in E, j \neq i} \left( \min_{c \in C_i, c' \in C_j} |c - c'| \right)
\end{equation}
A lower score indicates that elements are strictly aligned with each other's edges or centers.

\noindent\textbf{Readability (Rea $\downarrow$).}
We employ a computer vision-based approach to evaluate text legibility~\cite{zhou2022composition}. Unlike bounding-box approximations, we use the \textit{TreeWalker} API to extract precise rendering areas for all non-empty text nodes. 
For each text block $t_k$, we crop the corresponding region from the grayscale screenshot of the poster and compute the average gradient magnitude using the Sobel operator:
\begin{equation}
    G(t_k) = \frac{1}{|P_k|} \sum_{p \in P_k} \sqrt{G_x(p)^2 + G_y(p)^2}
\end{equation}
where $P_k$ is the set of pixels in the text block. The final score is the average gradient magnitude of all text blocks, normalized by the maximum possible Sobel magnitude (approx. 1442.5):
\begin{equation}
    \text{Rea} = \frac{1}{K} \sum_{k=1}^{K} \frac{G(t_k)}{1442.5}
\end{equation}
where $K$ denotes the total number of identified text blocks. Lower scores indicate smoother backgrounds (low-frequency regions) behind text, minimizing visual clutter and enhancing readability.

\noindent\textbf{CLIP Score (Clip $\uparrow$).} To evaluate the overall visual quality and style consistency, we compute the cosine similarity between the CLIP embeddings~\cite{radford2021learning} of the generated poster and the professional ground truth. This metric captures the model's ability to recover high-level design patterns and visual aesthetics.

\subsubsection{Subjective Metrics}

Automated metrics often fail to capture nuanced aesthetic qualities. We therefore employ a MLLM-based evaluation protocol.We utilize GPT-4.1-mini as a multi-modal judge to simulate human visual perception, evaluating the designs across four dimensions: \textbf{Text}, \textbf{Image}, \textbf{Layout}, and \textbf{Color}~\cite{ge2025autopresent}. Given the distinct data distributions between the Standard Benchmark and the Broad-Spectrum Benchmark (with the latter typically exhibiting greater complexity), we have tailored specific evaluation prompts for each scenario.

\begin{promptbox}{Evaluation Prompts for Standard Benchmark}
\setlength{\parindent}{0pt}
\small\ttfamily
Text. The text should be simple and clear to indicate the main point. Avoid too many texts and keep words concise. Use a consistent and readable font size, style, and color to attract attention. Avoid overlapping texts, as text collisions reduce readability. Avoid excessive or overly small text.

\vspace{1em}

Image. Use high-resolution, clear images. Penalize the design if images are blurry, pixelated, noisy, or have low resolution. Images must harmonize with the text. Do not treat decorative elements as images; score only the actual images. If the design does not contain any images, treat it as an intentional stylistic choice and give a neutral score instead of penalizing it.

\vspace{1em}

Layout. Elements should be aligned, do not overlap, and have sufficient margins to each other. The layout should not be too dense or too sparse. Since graphic design is typically a fixed, static medium rather than a dynamic, interactive digital screen, ensure that the layout adheres to graphic design styles and principles, avoiding web-style layouts such as buttons or floating panels.

\vspace{1em}

Color. Use high-contrast color especially between the text and the background. Avoid using high-glaring colors. Since graphic design is typically intended for print, the color palette should follow CMYK or CMYK-like visual logic, rather than relying on RGB screen-based effects.

\end{promptbox}

\begin{promptbox}{Evaluation Prompts for Broad-Spectrum Benchmark}
\setlength{\parindent}{0pt}
\small\ttfamily
Text. Text should be clear and readable, and must not be obscured. Penalize unreadable text or text that exceeds the boundaries. Text overlap and low legibility in peripheral areas and small-font regions should not be overlooked; these are also major issues within the text domain and should be penalized. Use a consistent and readable font size and style to attract attention. Since graphic design is typically intended for print, prohibit and punish web-style text, including but not limited to text with glowing neon effects. Text must be print-ready: sharp, high-contrast, and unobstructed.

\vspace{1em}

Image. Use high-resolution, clear images. Evaluate the image based on visual fidelity and the preservation of high-frequency details. Penalize the design if images are blurry, pixelated, noisy, or have low resolution. Penalize if the image looks like it was upscaled or heavily compressed, even if it is not overtly pixelated. Do not treat decorative elements as images; score only the actual images. If the design does not contain any images, treat it as an intentional stylistic choice and give a neutral score instead of penalizing it.

\vspace{1em}

Layout. Elements should be aligned, do not overlap, and have sufficient margins to each other. The overall layout density must be consistent; penalize layouts where some parts are too dense while others are too sparse. Assess the overall layout gestalt of the image. Graphic design is a static medium; it should not mimic the affordances of interactive digital products. If the design looks like a screenshot of a functional software, mobile app, or website, significantly lower the layout score. Penalize 'False Affordances': Harshly penalize any layout that creates an illusion of interactivity. Reject 'Containerization': Web layouts use rigid boxes/containers to manage dynamic content (responsiveness). Good graphic design uses composition and whitespace. Punish layouts that rely on 'boxing' content unnecessarily.

\vspace{1em}

Color. Use high-contrast color especially between the text and the background. Avoid using high-glaring colors. Penalize 'Emissive' Aesthetics: Downgrade designs where colors appear to be glowing, backlit, or self-luminous. Graphic design relies on pigment and ink, which are subtractive and reflective. If the image looks like it would require a screen to exist, score it low. Avoid overuse of high-brightness, overly saturated, or neon-like colors that are typically intended for screen displays.

\end{promptbox}

\begin{promptbox}{System Prompt for Evaluation (Same for Both benchmarks)}
\setlength{\parindent}{0pt}
\small\ttfamily
Please evaluate the graphic design based on the following criteria.

\vspace{1em}

Give an integer score between 0 - 5, higher scores means the criteria is better met.
First, respond with a score; Then, provide your justification for the score in a natural language sentence (one sentence is enough). Your response should look like this: '4. The graphic design has clear texts good color matching.'
Only evaluate the graphic design based on the specified criteria, and no other aspects. Give scores across the full spectrum (1-5) instead of only good ones (3-5).

\end{promptbox}

For MLLM-based evaluation, we use a 5-point rating scale and convert the raw scores to a 100-point scale by multiplying them by 20.

\subsubsection{Human Evaluation}
To mitigate cognitive overload on human evaluators and maintain the reliability of the rankings, we distilled the four subjective evaluation dimensions into two comprehensive metrics: \textit{Structure Rationality} and \textit{Visual Aesthetics}. Their detailed definitions are as follows:

\begin{itemize}
    \item \textbf{Structure Rationality:} Evaluates two main aspects: 1) \textit{Text Readability} (ensuring text is clear, legible, and structurally complete); and 2) \textit{Layout Coherence} (ensuring elements are properly aligned without visual clutter or overcrowding).
    \item \textbf{Visual Aesthetics:} Evaluates the overall artistic appeal, specifically ranking outputs based on \textit{Image Quality} and \textit{Color Harmony}.
\end{itemize}

\subsection{Prompts in the Plan-Implementation-Reflect Pipeline}
\label{apd:PIR_prompts}
\begin{promptbox}{Prompts of Semantic Planner}
\begin{lstlisting}
You are a master of 2D graphic design. You are skilled in planning 2D design, adept at providing design concepts and layout thought, and capable of generating corresponding grouping plans, image prompts and text content based on the layout thought.  

Workflow:  

1. Provide the `layout_thought`, enclosed in <layout_thought>...</layout_thought>. As detailed as possible, including the layout structure and any specific elements (layers).

2. Provide the `grouping`, enclosed in <grouping>...</grouping>. It should be a JSON array that groups related layers together. Each group must be expressed as a JSON object with three fields:
- `group_id`: a unique identifier string like 'G1', 'G2'...
- `children`: a list of layer_ids (from the `layout_thought` you just generated) that belong to this group.
- `theme`: a short description (2-6 words) summarizing the group's purpose (e.g., 'menu item', 'header block').
The `grouping` should help later stages bind text and image elements correctly.
If an element is standalone and not obviously related, it can form its own group.
The grouping must appear right after the layout_thought, and will guide the subsequent image and text generation.

3. Provide the image generation prompts, enclosed in <image_generator>...</image_generator>, for example:  

<image_generator>
[
{"layer_id": 0, "layer_prompt": "prompt0"},
{"layer_id": 1, "layer_prompt": "prompt1"}
]
</image_generator>

4. Provide the text element design, enclosed in <generate_text>...</generate_text>. Example:  

<generate_text>
[
{
    "layer_id": 6,
    "type": "TextElement",
    "width": 302.1794738769531,
    "height": 31.327075958251953,
    "opacity": 1.0,
    "text": "Big Fall Volunteer",
    "font": "Abril Fatface",
    "font_size": 31.39527130126953,
    "text_align": "center",
    "angle": 0.0,
    "capitalize": false,
    "line_height": 1.0,
    "letter_spacing": 0.9849796295166016
},
{
    "layer_id": 7,
    "type": "TextElement",
    "width": 322.0,
    "height": 67.89791107177734,
    "opacity": 1.0,
    "text": " Cleanup",
    "font": "Abril Fatface",
    "font_size": 68.0,
    "text_align": "center",
    "angle": 0.0,
    "capitalize": false,
    "line_height": 1.0,
    "letter_spacing": 0.0
}
]
</generate_text>

Important:  
- <layout_thought>...</layout_thought>, <grouping>...</grouping>, <image_generator>...</image_generator>, and <generate_text>...</generate_text> are mandatory and must appear exactly once. 

\end{lstlisting}
\end{promptbox}

\begin{promptbox}{Prompts of Composer}
\begin{lstlisting}
You are a **2D graphic design master**, highly skilled in generating a wide range of visual designs, especially **posters, advertisements, and promotional pages in HTML format**.  

When creating the final output, you must follow and consider these key elements:  
- The **user's input**  
- The **intermediate reasoning process (layout_thought)**  
- Content generated by various tools, including **text** and **images** (insert image URLs directly into the HTML)  

**Strict requirement:**
- The final HTML **must include a main container with the class** .poster.
- All design content (background, images, text, decorations) must be placed **inside** .poster.

Your output should be a 2D graphic design in the form of HTML. Just return the HTML. Do not return anything else except this. Do not place the HTML in the code block.
\end{lstlisting}
\end{promptbox}

\begin{promptbox}{Prompts of Refiner}
\begin{lstlisting}
You are a professional graphic designer.
Based on the original HTML, its rendered image, and the refined reference image, improve the HTML code:
- Enhance visual aesthetics (colors, typography, spacing, alignment).
- Ensure layout balance and professional design.
- Preserve semantic correctness and responsiveness.
Return only the improved HTML. Do not place the HTML in the code block.

**Strict requirement:**
- The HTML **must include a main container with the class** .poster.
- All design content (background, images, text, decorations) must be placed **inside** .poster."

\end{lstlisting}
\end{promptbox}

\subsection{Semantic Planner Training Data Format}
\label{apd:data}

\begin{promptbox}{Example for Semantic Planner Training Data}
\setlength{\parindent}{0pt}
\raggedright
\small\ttfamily

<user\_input>

A promotional poster for a mountain biking championship.

</user\_input>
\vspace{1em}

<layout\_thought>

This layout should be dynamic and engaging, utilizing a central alignment strategy with overlapping elements to create depth and visual interest. The design should emphasize the action and energy of the event, with bold typography and a vibrant color palette to capture attention. Each layer should serve a specific purpose to enhance the hierarchy of information and overall aesthetic.

The background layer (Layer 0) should be a solid black to create a sleek and professional foundation. This dark backdrop will allow the vibrant colors and white elements to stand out clearly, ensuring high contrast and visual clarity.

Next, the diagonal blue and orange shapes (Layers 1 and 2) should be placed to create a dynamic composition. The blue shape (Layer 1) should be positioned slightly to the left, tilted at an angle to add motion and energy to the layout. Its large size should make it a dominant background element. The orange shape (Layer 2) should overlap with the blue shape, tilted in the opposite direction to create a sense of balance and vibrancy. These shapes should have full opacity and vibrant colors to enhance the visual appeal.

The central focal point should be the illustrated figure of the cyclist (Layer 3). This element should be positioned slightly off-center to the left, dynamically interacting with the diagonal background shapes. The cyclist's forward motion should guide the viewer's eye toward the right side of the layout, where the main event details are located. The cyclist's colors should harmonize with the overall palette while maintaining enough contrast to ensure prominence.

The headline text "FREERIDE" (Layer 4) should be placed on the right side of the layout, horizontally aligned with the cyclist's motion. It should use a bold, white typeface to grab attention immediately. The large font size ensures that it is the most prominent text element, emphasizing the event's name.

Above "FREERIDE," the date "OCTOBER 20" (Layer 5) should be placed in smaller text, aligned to the right to maintain symmetry. Its white color ensures it stands out against the dark background, while its smaller size ensures it does not compete with the headline.

Below "FREERIDE," the subheading "CHAMPIONSHIP" (Layer 6) should be positioned. It should use a smaller, bold typeface to indicate the event's significance. Its placement below the main headline creates a logical reading flow.

The supporting text "8AM-5PM / SUNDAY" (Layer 7) should be placed further down, aligned to the right. Its smaller size and bold weight ensure it is noticeable but does not overpower the other text elements. It provides essential event details and complements the overall composition.

Finally, the website URL "www.bmx.com" (Layer 8) should be placed at the bottom right corner. Its smaller size and white color ensure it remains legible but unobtrusive. This positioning provides closure to the layout, directing the viewer's attention to the call-to-action after the other elements have been absorbed.

This layout should balance action, energy, and clarity, ensuring that the viewer's attention is drawn to the most critical elements first while maintaining an overall cohesive design.

</layout\_thought>
\vspace{1em}

<grouping>

[{"group\_id": "G1", "children": [0], "theme": "background color"}, {"group\_id": "G2", "children": [1, 2], "theme": "diagonal background shapes"}, {"group\_id": "G3", "children": [3], "theme": "illustration of cyclist"}, {"group\_id": "G4", "children": [4, 5, 6, 7, 8], "theme": "event information text"}]

</grouping>
\vspace{1em}

<image\_generator>

[{"layer\_id": 0, "layer\_prompt": "Create a solid rectangular background layer, 1920 × 1080 px, filled entirely with pure black (hex \#000000), 100 \% opacity, no gradients, patterns, or textures."}, {"layer\_id": 1, "layer\_prompt": "Create a 1246 × 1246 px layer containing one solid, fully opaque sky-blue (\#46B9E9) parallelogram band that runs diagonally from the upper left toward the lower right at roughly a 35–40° angle, spanning the full width of the canvas; every area outside the band is completely transparent"}, {"layer\_id": 2, "layer\_prompt": "A single, bold diagonal stripe forming a long parallelogram that runs from the upper-left to the lower-right of the canvas, filled with solid orange (\#F48729) at full opacity; every part of the canvas outside this stripe is completely transparent"}, {"layer\_id": 3, "layer\_prompt": "Flat vector illustration, 1101 × 902 px, opacity 100 \%.  
Content: a dynamic low-angle view of a BMX rider kicking up a cloud of dust on a straight two-lane road that stretches into the distance.  
• Rider: crouched, wearing a red jersey with white side panels, black shorts, dark blue knee-length socks, grey/black cycling shoes, grey helmet, black gloves, and a flowing black cape.  
• Bike: vivid orange BMX with thin frame, large black wheel rims, silver spokes, black saddle, and grey handle grips.  
• Road: simple flat grey surface with a dashed white centre line and matching white edge lines.  
• Dust: stylized swirling cloud of light grey with a darker grey core, partially obscuring the rider and front wheel.  
Transparent areas: everything outside the rider, bike, road, and dust cloud is fully transparent."}]

</image\_generator>
\vspace{1em}

<generate\_text>

[{"layer\_id": 4, "type": "TextElement", "width": 761.0, "height": 142.0, "opacity": 1.0, "text": "FREERIDE", "font": "Knewave", "font\_size": 142.0, "text\_align": "left", "angle": 0.0, "capitalize": false, "line\_height": 1.0, "letter\_spacing": 0.0}, {"layer\_id": 5, "type": "TextElement", "width": 406.0, "height": 49.0, "opacity": 1.0, "text": "OCTOBER 20", "font": "Montserrat", "font\_size": 49.0, "text\_align": "left", "angle": 0.0, "capitalize": false, "line\_height": 1.0, "letter\_spacing": 0.0}, {"layer\_id": 6, "type": "TextElement", "width": 413.0, "height": 40.0, "opacity": 1.0, "text": "CHAMPIONSHIP", "font": "Montserrat", "font\_size": 40.0, "text\_align": "left", "angle": 0.0, "capitalize": false, "line\_height": 1.0, "letter\_spacing": 0.0}, {"layer\_id": 7, "type": "TextElement", "width": 352.0, "height": 32.0, "opacity": 1.0, "text": "8AM-5PM / SUNDAY", "font": "Montserrat", "font\_size": 32.0, "text\_align": "left", "angle": 0.0, "capitalize": false, "line\_height": 1.0, "letter\_spacing": 0.0}, {"layer\_id": 8, "type": "TextElement", "width": 207.0, "height": 25.0, "opacity": 1.0, "text": "www.bmx.com", "font": "Montserrat", "font\_size": 25.0, "text\_align": "left", "angle": 0.0, "capitalize": false, "line\_height": 1.0, "letter\_spacing": 0.0}]

</generate\_text>

\end{promptbox}


\subsection{Inference Time and Computational Cost}
\label{apd:inference_time}

The end-to-end generation process of our proposed DesignAsCode framework requires multiple sequential operations, including semantic planning, visual asset generation or retrieval, HTML/CSS composition, and visual-aware reflection. On average, generating a complete, high-fidelity, and fully editable graphic design takes approximately 1 to 5 minutes, depending on the complexity of the layout and the number of image layers to be generated.

While this inference time is naturally higher than that of traditional, single-pass text-to-image models, this computational cost is highly practical and negligible when contextualized within the broader landscape of modern Large Language Model (LLM) agentic workflows. In the current era of autonomous AI systems, multi-step generation pipelines routinely exchange longer processing times for superior accuracy and structural control. For instance, in academic literature, automated LLM agents designed for project-level code execution report average runtimes of 74 minutes per task, a computational cost explicitly deemed ``reasonable'' given the complexity of the autonomous workflow~\cite{bouzenia2025you}. Similarly, for reasoning and retrieval tasks, advanced agent systems routinely exhibit latencies ranging from several minutes to over 10 minutes per query~\cite{zhang2025optimizing}.

Given that our \textit{Plan-Implement-Reflect} pipeline requires zero human intervention during the generation process and yields professional-grade, fully editable code representations, a multi-minute inference time is a standard and highly acceptable trade-off. Furthermore, this fully automated nature makes DesignAsCode exceptionally well-suited for asynchronous execution and large-scale batch personalization tasks, where manual drafting would otherwise consume significantly more human hours.

\subsection{Details of Attribute-Level Editability Controlled by CSS Code}
\label{theme_switch}

\begin{figure}[H]
  \centering
  \includegraphics[
    width=\textwidth,
    trim=20 250 60 50,
    clip
  ]{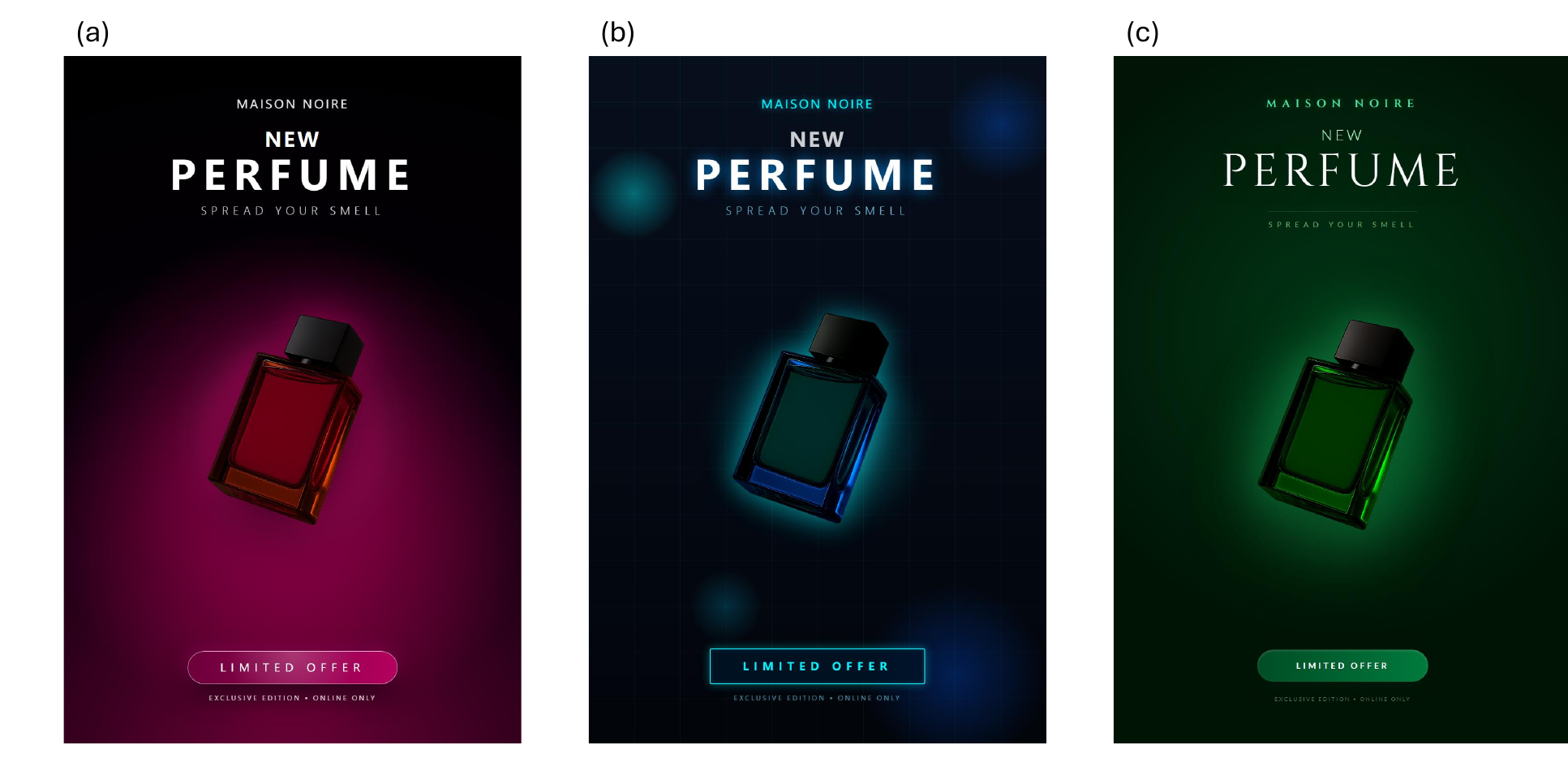}
  \caption{Code-controlled theme switch corresponding to Appendix~\ref{theme_switch}. From this figure, we can also observe that CSS enhances the product’s highlight effects.}
  \label{fig:perfume_compare}
\end{figure}

Figure~\ref{fig:perfume_compare} and the specific CSS code snippets below demonstrate DesignAsCode's powerful, code-controlled Attribute-Level Editability through a concrete example.

\begin{codebox}{Original CSS Styling for the Product Image (Figure~\ref{fig:perfume_compare}a)}
.product-image {
        width: 520px;
        max-width: min(60vw, 520px);
        display: block;
        filter:
          drop-shadow(0 42px 90px rgba(0, 0, 0, 1))
          drop-shadow(0 0 42px rgba(255, 0, 140, 0.7));
        transform: translateY(-8px);
      }
\end{codebox}

\begin{codebox}{Core CSS Code for the Blue Color Transformation (Figure~\ref{fig:perfume_compare}b)}
.product-image {
        width: 520px;
        max-width: min(60vw, 520px);
        display: block;
        transform: translateY(-8px);
        
        /* Core style transfer code:
           1. hue-rotate(190deg): Rotates the hue of the red perfume bottle to turn it cyan-blue.
           2. brightness/contrast: Increases brightness and contrast to mimic a luminous object.
           3. drop-shadow: Adds a strong cyan outer glow.
        */
        filter: 
          hue-rotate(190deg) 
          saturate(150%) 
          brightness(1.1) 
          contrast(1.2) 
          drop-shadow(0 0 30px rgba(0, 243, 255, 0.5));
      }
\end{codebox}

\begin{codebox}{Core CSS Code for the Green Color Transformation (Figure~\ref{fig:perfume_compare}c)}
.product-image {
        width: 520px;
        max-width: min(60vw, 520px);
        display: block;
        transform: translateY(-8px);
        
      /* --- Core Modification Area --- */
        filter: 
          hue-rotate(110deg) 
          saturate(1.6) 
          contrast(1.1) 
          brightness(1.3)
          drop-shadow(0 40px 80px rgba(0, 40, 10, 0.8))
          drop-shadow(0 0 40px rgba(50, 255, 150, 0.4)); /* Add outer glow */
      }
\end{codebox}

\subsection{Discussion on Model Dependency and Open-Source Applicability}
\label{appendix:model_dependency}

While our \textit{DesignAsCode} framework establishes a new state-of-the-art in editable graphic design generation, we acknowledge that our current implementation relies on closed-source commercial models (e.g., GPT-5 as the layout Composer and Refiner, and GPT-image-1 for image editing). We would like to clarify the rationale behind this choice and the future trajectory of our framework.

\textbf{Proof of Concept for a Novel Paradigm.} 
The primary contribution of this work is the programmatic HTML/CSS synthesis paradigm and the novel Plan-Implement-Reflect pipeline. To rigorously evaluate the theoretical upper bound of this paradigm and to demonstrate that generating complex, unconstrained HTML/CSS code natively resolves visual dissonance, we required the most capable multimodal models available. Currently, generating rich, nested web code interweaved with visual feedback is a task where state-of-the-art proprietary models still hold a noticeable edge over existing open-weights alternatives.

\textbf{Framework Modularity and Model Agnosticism.} 
Crucially, our pipeline is strictly modular and model-agnostic. It does not rely on any proprietary internal representations or closed APIs beyond standard text/image input-output. As open-source Multimodal Large Language Models (MLLMs) rapidly close the capability gap, any module in our pipeline can be seamlessly swapped out as a ``plug-and-play'' component.

\textbf{Demonstrated Open-Source Feasibility.} 
We have already taken significant steps toward a fully open-weights pipeline. As detailed in our methodology, the Semantic Planner---which handles the critical and complex task of layout reasoning and hierarchical grouping---is successfully powered by a lightweight, open-source LLM (Qwen3-8B) fine-tuned on our distilled dataset. This demonstrates that the high-level design logic required by our framework can be efficiently internalized by smaller open models. We anticipate that as open-source visual-coding models mature, the Composer and Refiner modules can be similarly localized without sacrificing the high visual fidelity and structural editability our method achieves.

\section{Additional Results}

\subsection{More Qualitative Comparisons between DesignAsCode and Baselines}
\begin{figure}[H]
  \centering
  \includegraphics[
    width=\textwidth,
    trim=0 100 0 0,
    clip
  ]{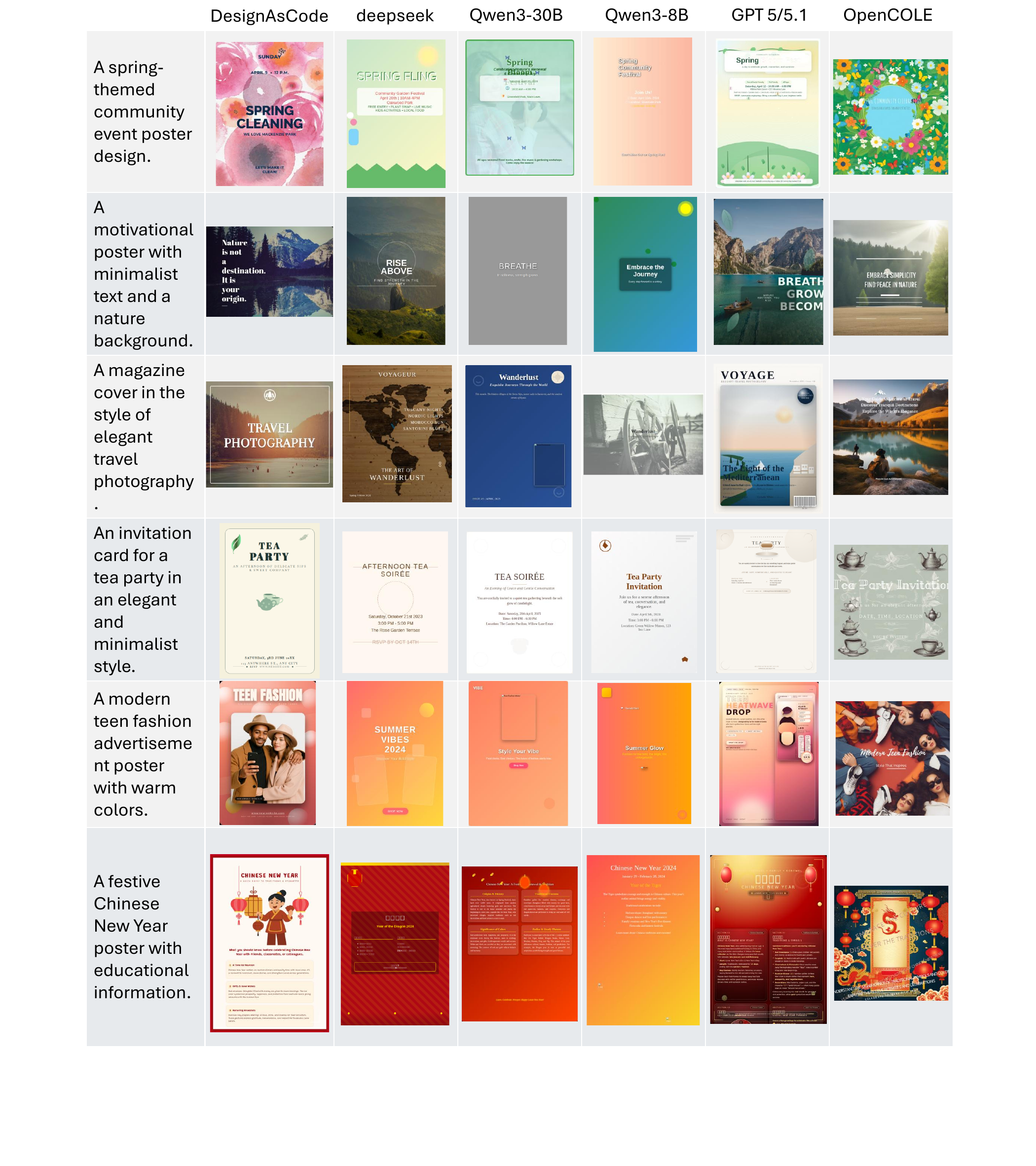}
  \caption{More comparisons between DesignAsCode and baselines. The leftmost column displays the input text prompts. The top three rows demonstrate the results on the Crello test set, while the bottom three rows show the results on the Broad test set.}
  \label{fig:quality_2}
\end{figure}

Figure~\ref{fig:quality_2} presents additional qualitative comparative results between DesignAsCode and baselines.

\subsection{More Quantitative Comparisons between DesignAsCode and Baselines}
\label{apd:more_quantitative_comparisons}
To further compare DesignAsCode with prior agent-based graphic design generation methods, we reproduce BannerAgency~\cite{wang2025banneragency} and evaluate both methods on the Crello benchmark. As shown in Table~\ref{tab:banneragency_comparison}, DesignAsCode achieves better performance on Validity, Readability, and CLIP Score, while obtaining substantially higher scores across all four subjective dimensions. BannerAgency attains a slightly lower Alignment score, but its fixed logo--background--foreground schema limits its applicability to more diverse design tasks. In contrast, DesignAsCode dynamically constructs task-specific structures from a single prompt.

\subsection{Quantitative Evaluation of Editability}
\label{apd:editability}

\begin{table}[t]
  \caption{Comparison with BannerAgency on Crello. DesignAsCode achieves stronger overall performance across most objective metrics and all subjective metrics.}
  \label{tab:banneragency_comparison}
  \centering
  \small
  \setlength{\tabcolsep}{3pt}
  \begin{tabular*}{\textwidth}{@{\extracolsep{\fill}}lcccccccc@{}}
    \toprule
    \multirow{2}{*}{Method}
      & \multicolumn{4}{c}{Objective Metrics}
      & \multicolumn{4}{c}{Subjective Metrics} \\
    \cmidrule(lr){2-5}\cmidrule(lr){6-9}
      & Val $\uparrow$ & Ali $\downarrow$ & Rea $\downarrow$ & CLIP $\uparrow$
      & Text $\uparrow$ & Image $\uparrow$ & Layout $\uparrow$ & Color $\uparrow$ \\
    \midrule
    BannerAgency
      & 0.9088 & \textbf{0.0003} & 0.1025 & 0.5760
      & 50.15 & 43.19 & 41.03 & 59.05 \\
    \textbf{DesignAsCode (Ours)}
      & \textbf{0.9521} & 0.0008 & \textbf{0.0849} & \textbf{0.6287}
      & \textbf{74.06} & \textbf{65.86} & \textbf{63.67} & \textbf{68.42} \\
    \bottomrule
  \end{tabular*}
\end{table}

We quantitatively evaluate the editability of the generated designs on the Crello test set from two complementary perspectives. First, we conduct an \textit{editability audit} to verify whether common design attributes remain explicitly accessible in the generated HTML/CSS. As shown in Table~\ref{tab:editability_audit}, all evaluated designs expose editable text, images, colors, fonts, positions, and element groups, confirming that our method preserves structured design components rather than flattening them into raster images.

\begin{table}[t]
  \caption{Editability audit on generated Crello designs. The generated HTML/CSS explicitly exposes common editable design attributes.}
  \label{tab:editability_audit}
  \centering
  \small
  \begin{tabular*}{\textwidth}{@{\extracolsep{\fill}}lcccccc@{}}
    \toprule
    Attribute & Text & Image & Color & Font & Position & Group \\
    \midrule
    Success Rate & 100.0\% & 100.0\% & 100.0\% & 100.0\% & 100.0\% & 100.0\% \\
    \bottomrule
  \end{tabular*}
\end{table}

Second, we perform an \textit{edit-preservation analysis} using three common programmatic edits: text replacement, color modification, and font-size adjustment. After each edit, we re-render the design and measure changes in Validity, Alignment, and Readability relative to the original rendering. Because these edits are intended to test structural preservation rather than improve visual aesthetics, subjective evaluation is not included. As shown in Table~\ref{tab:edit_preservation}, all edits succeed, while the objective metrics change only negligibly. These results demonstrate that DesignAsCode supports reliable local editing without disrupting the surrounding layout.
Qualitative examples of CSS-level theme editing are provided in Supplementary Material~\ref{theme_switch}.

\begin{table}[t]
  \caption{Edit-preservation analysis on generated Crello designs. Common local edits preserve layout quality with negligible metric changes.}
  \label{tab:edit_preservation}
  \centering
  \small
  \begin{tabular*}{\textwidth}{@{\extracolsep{\fill}}lcccc@{}}
    \toprule
    Edit Type & Success Rate & $\Delta$Val & $\Delta$Ali & $\Delta$Rea \\
    \midrule
    Text Replacement & 100.0\% & 0.0147 & $-0.0000$ & 0.0005 \\
    Color Modification & 100.0\% & 0.0000 & $-0.0000$ & $-0.0000$ \\
    Font-Size Adjustment & 100.0\% & 0.0000 & 0.0000 & 0.0002 \\
    \bottomrule
  \end{tabular*}
\end{table}

\subsection{More Results Generated by DesignAsCode}
\begin{figure}[H]
  \centering
  \includegraphics[
    width=\textwidth,
    trim=0 170 0 0,
    clip
  ]{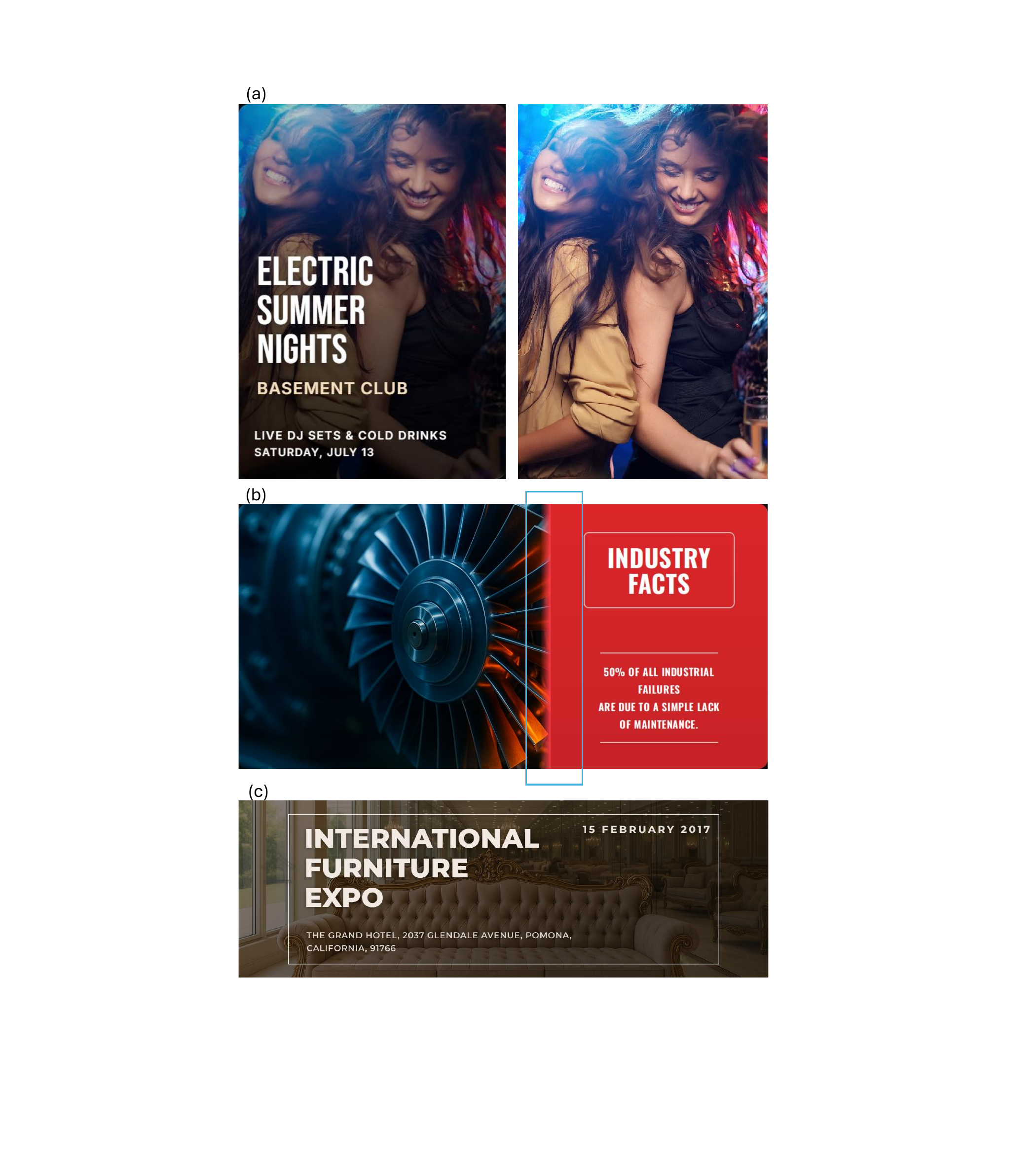}
  \caption{Decorative effects implemented via HTML/CSS by DesignAsCode. (a) Left: A design using blend modes, contrasting with the original background image on the right. (b): A design featuring a glow effect (highlighted in the blue box). (c): A design utilizing shadows and gradients, where the background treatment effectively highlights the text theme.}
  \label{fig:html_effect}
\end{figure}

\begin{figure}[H]
  \centering
  \includegraphics[
    width=\textwidth,
    trim=0 50 0 0,
    clip
  ]{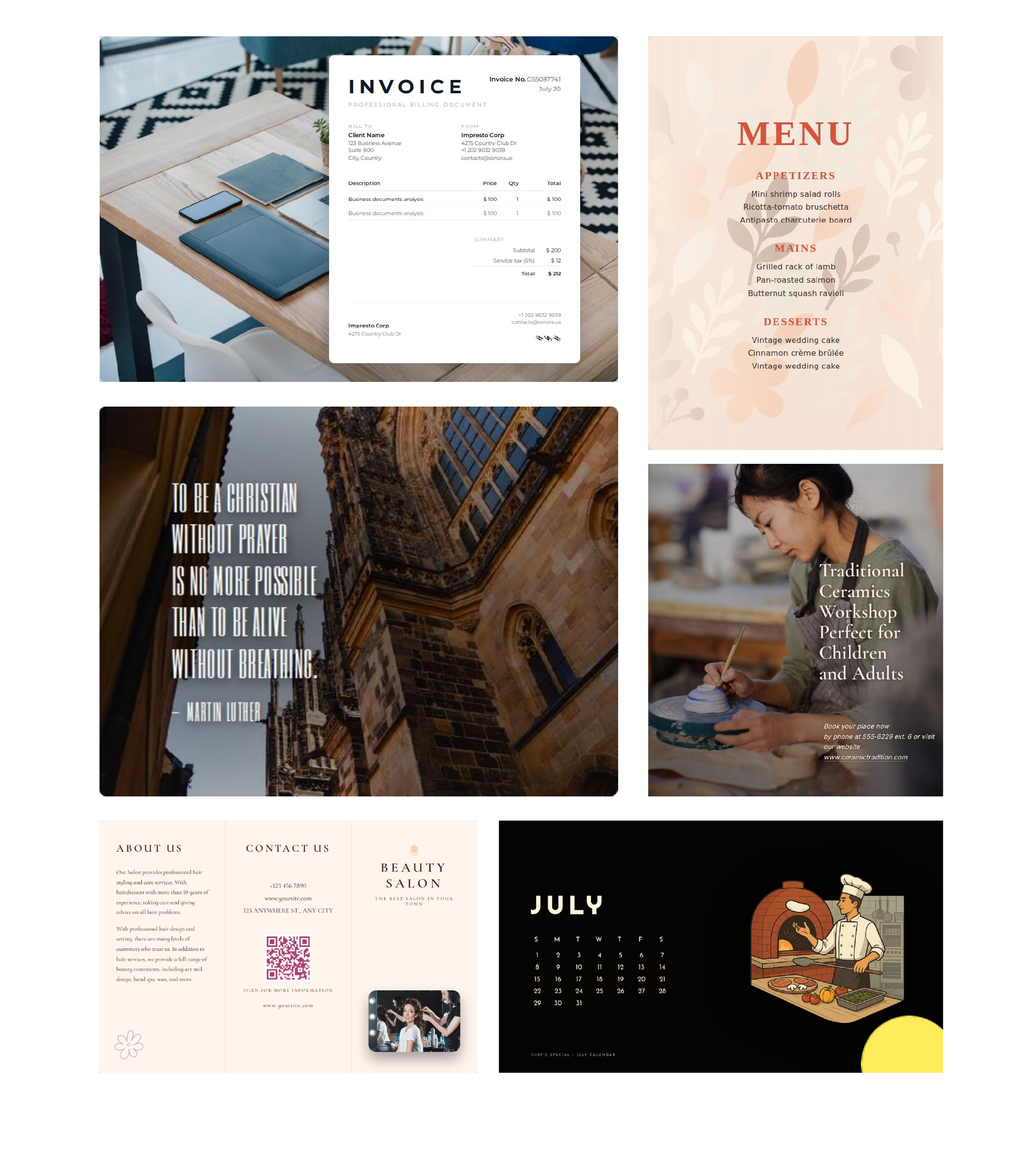}
  \caption{More results generated by DesignAsCode.}
  \label{fig:more_result}
\end{figure}

\begin{figure}[H]
  \centering
  \includegraphics[
    width=\textwidth,
    trim=0 50 0 0,
    clip
  ]{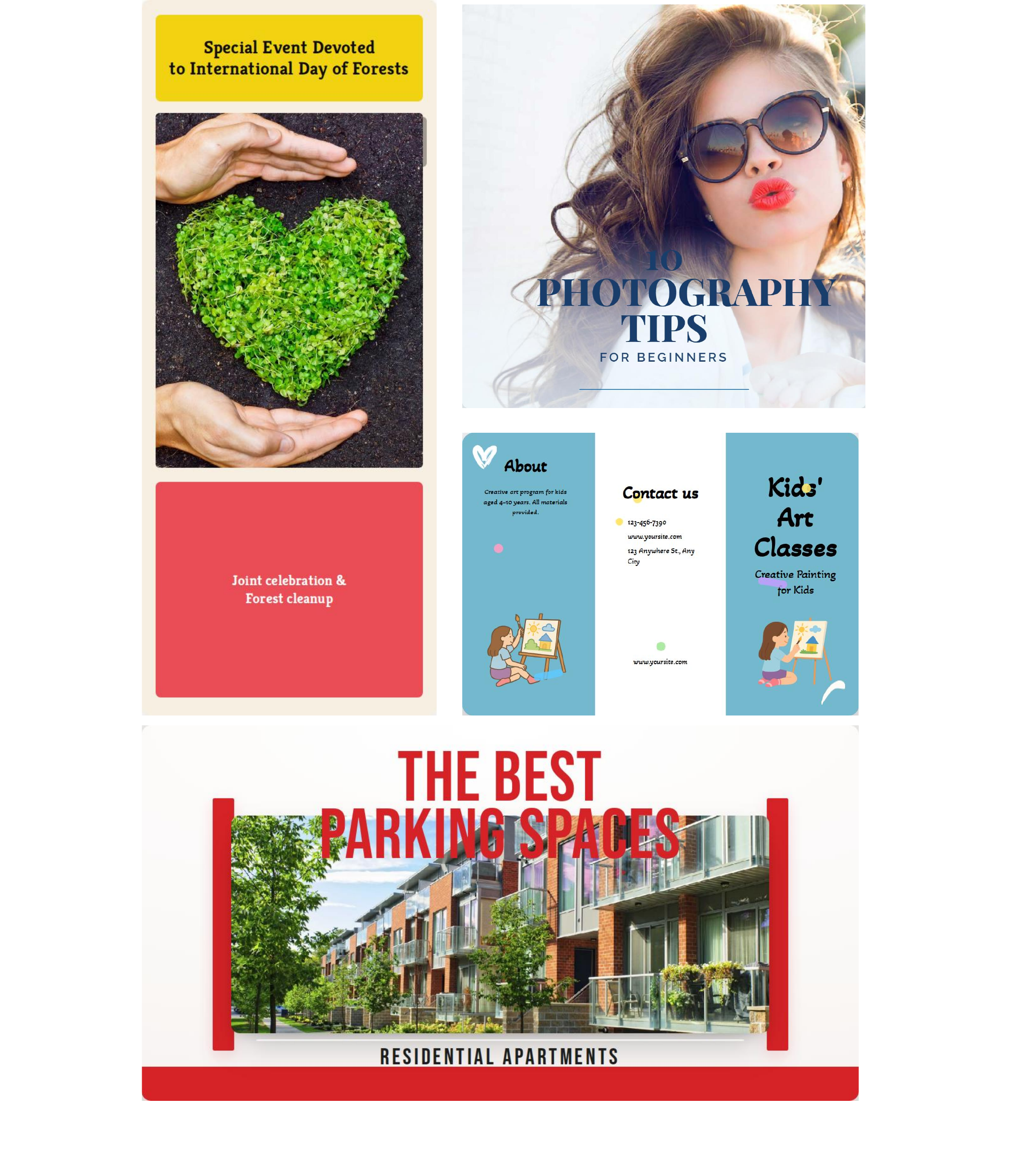}
  \caption{More results generated by DesignAsCode.}
  \label{fig:more_result_2}
\end{figure}

\begin{figure}[H]
  \centering
  \includegraphics[
    width=\textwidth,
    trim=0 50 0 0,
    clip
  ]{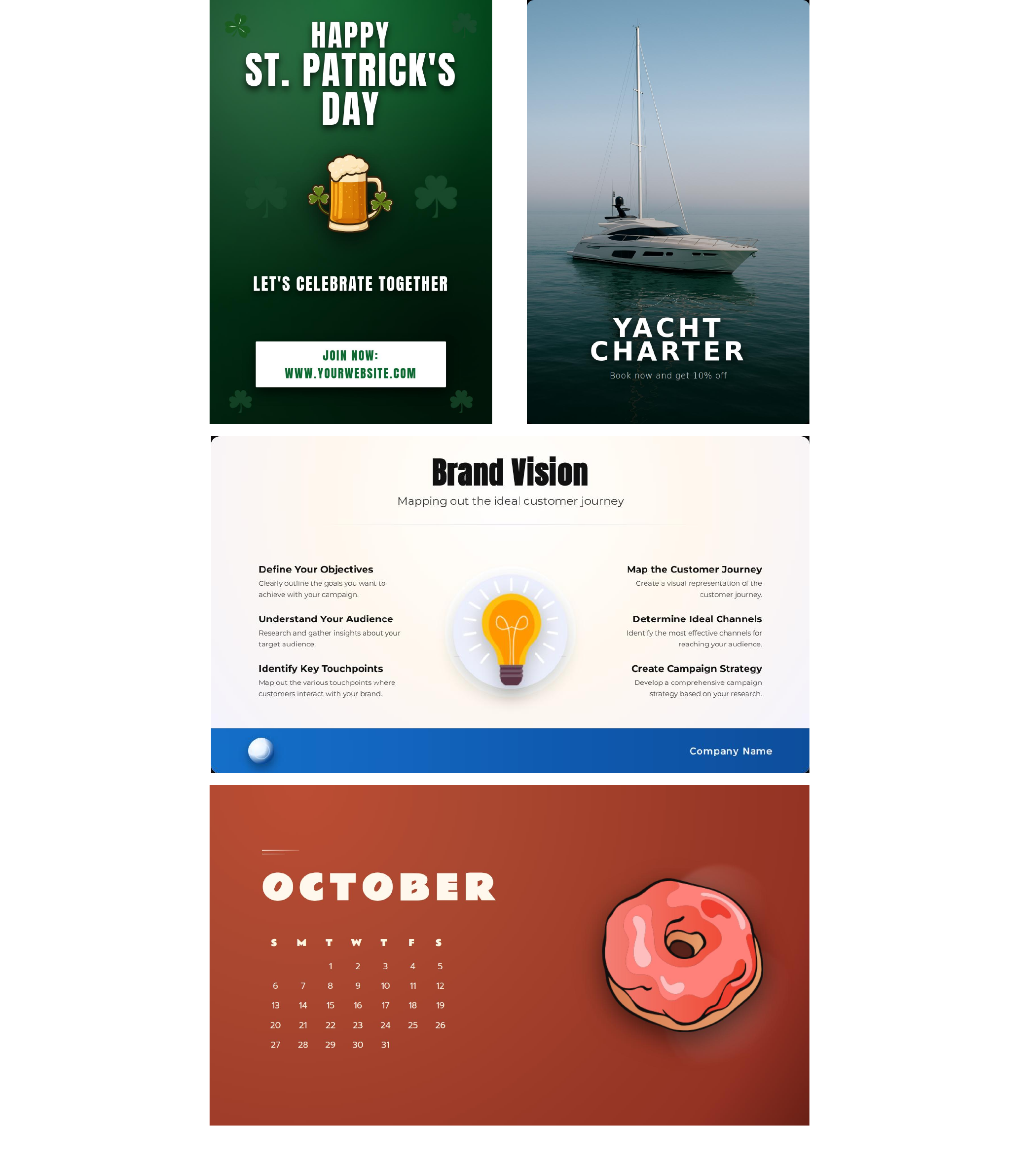}
  \caption{More results generated by DesignAsCode.}
  \label{fig:more_result_3}
\end{figure}

Figure~\ref{fig:html_effect} presents more detailed decorative effects implemented via HTML/CSS by DesignAsCode. Figure~\ref{fig:more_result}, \ref{fig:more_result_2} and \ref{fig:more_result_3} present more results generated by DesignAsCode.

\subsection{A Complete HTML Design and Matching Rendering Example}
\label{code:team}
\begin{figure}[H]
  \centering
  \includegraphics[
    width=\textwidth,
    trim=0 0 0 200,
    clip
  ]{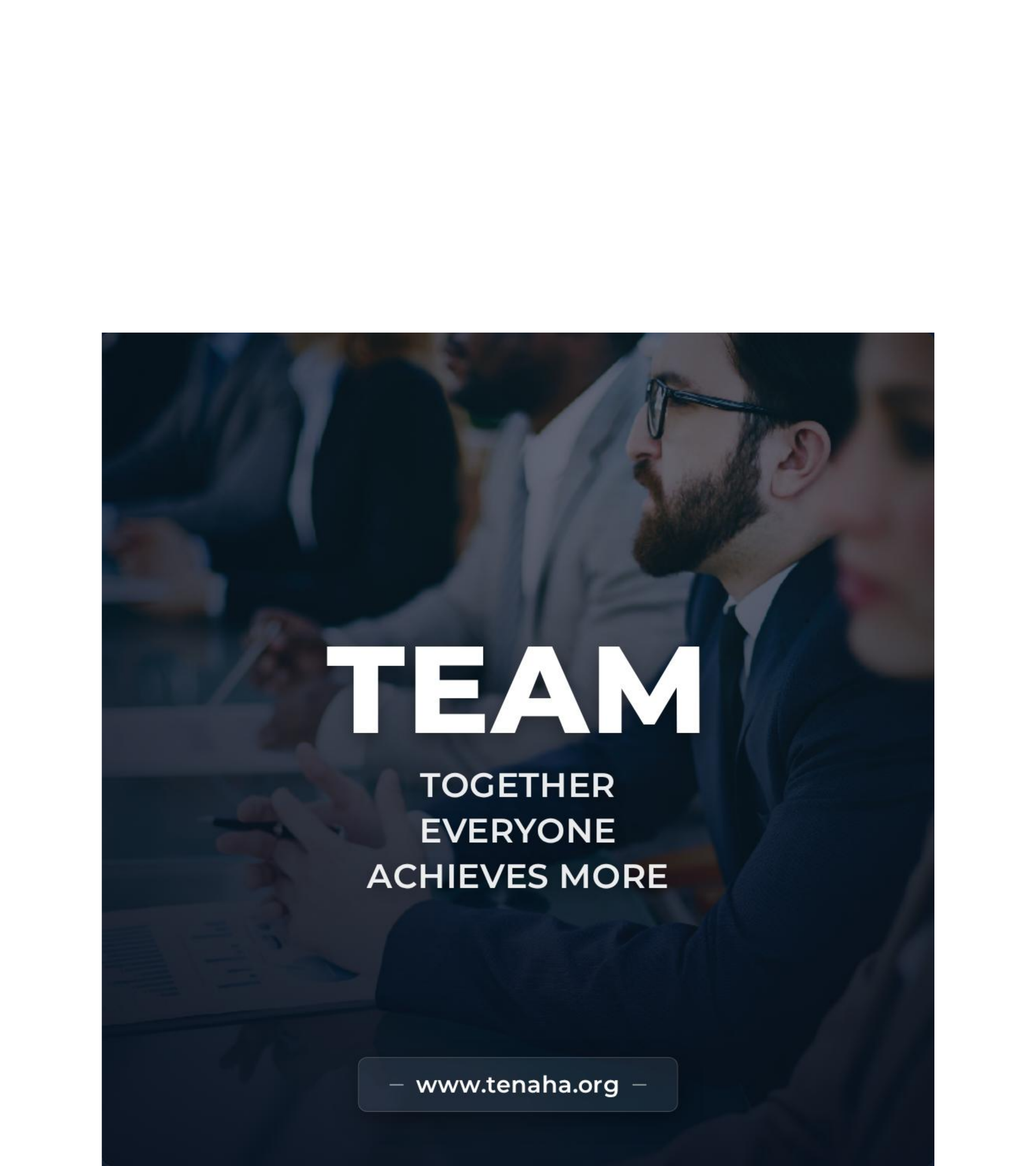}
  \caption{Rendering corresponding to Appendix~\ref{code:team}}
  \label{fig:corresponding}
\end{figure}

This section presents a complete HTML/CSS design generated by DesignAsCode, with its corresponding rendered image illustrated in Figure~\ref{fig:corresponding}.
\begin{codebox}{HTML/CSS Code for Figure~\ref{fig:corresponding}}
<!DOCTYPE html>
<html lang="en">
<head>
  <meta charset="UTF-8"/>
  <meta name="viewport" content="width=device-width, initial-scale=1.0"/>
  <title>Corporate Marketing Team Meeting Poster</title>
  <link href="https://fonts.googleapis.com/css2?family=Montserrat:wght@500;600;800&display=swap" rel="stylesheet">
  <style>
    :root { color-scheme: dark; }
    body {
      margin: 0;
      background: #0d1c32;
      display: grid;
      place-items: center;
      min-height: 100vh;
    }
    .poster {
      position: relative;
      width: min(92vw, 1080px);
      height: min(92vw, 1080px);
      overflow: hidden;
      font-family: 'Montserrat', Arial, sans-serif;
      color: #fff;
      border-radius: 12px;
    }
    .poster .bg,
    .poster .texture {
      position: absolute;
      inset: 0;
      width: 100%;
      height: 100%;
      object-fit: cover;
    }
    .poster .texture {
      z-index: 2;
      mix-blend-mode: multiply;
      opacity: .65;
      pointer-events: none;
    }
    /* Gradient + vignette for legibility */
    .poster::after {
      content: "";
      position: absolute;
      inset: 0;
      z-index: 3;
      pointer-events: none;
      background:
        radial-gradient(120% 120% at 75% 25%, rgba(6,15,29,.0) 0%, rgba(6,15,29,.45) 58%, rgba(6,15,29,.78) 100%),
        linear-gradient(to bottom, rgba(10,18,30,0) 15%, rgba(10,18,30,.35) 55%, rgba(10,18,30,.75) 100%);
    }

    .content {
      position: absolute;
      z-index: 5;
      left: 50%;
      top: 50%;
      transform: translate(-50%, -44%);
      width: 78%;
      max-width: 760px;
      text-align: center;
    }
    .title {
      margin: 0 0 clamp(12px, 2.5vw, 22px);
      font-weight: 800;
      text-transform: uppercase;
      letter-spacing: 2px;
      line-height: .92;
      font-size: clamp(86px, 12.5vw, 160px);
      text-shadow: 0 10px 26px rgba(0,0,0,.35);
    }
    .subtitle {
      margin: 0 auto;
      max-width: 520px;
      font-weight: 600;
      letter-spacing: .6px;
      line-height: 1.34;
      font-size: clamp(26px, 3.6vw, 44px);
      color: rgba(255,255,255,.92);
      text-transform: uppercase;
      text-shadow: 0 6px 18px rgba(0,0,0,.28);
    }

    .url {
      position: absolute;
      z-index: 6;
      left: 50%;
      bottom: clamp(28px, 5vw, 70px);
      transform: translateX(-50%);
      display: inline-flex;
      align-items: center;
      justify-content: center;
      padding: clamp(10px, 1.8vw, 16px) clamp(16px, 2.6vw, 24px);
      font-weight: 600;
      font-size: clamp(18px, 2.4vw, 30px);
      letter-spacing: .4px;
      color: #fff;
      text-decoration: none;
      background: rgba(255,255,255,.08);
      border: 1px solid rgba(255,255,255,.18);
      border-radius: 12px;
      backdrop-filter: saturate(120%) blur(2px);
      box-shadow: 0 10px 30px rgba(0,0,0,.35);
    }
    .url::before,
    .url::after {
      content: "";
      display: block;
      width: clamp(12px, 1.8vw, 18px);
      height: 0;
      border-top: 2px solid rgba(255,255,255,.6);
      margin: 0 clamp(10px, 1.4vw, 16px);
      opacity: .8;
    }

    /* Safe-area padding on small screens */
    @media (max-width: 560px) {
      .content { transform: translate(-50%, -46%); width: 86%; }
      .poster { border-radius: 8px; }
    }
  </style>
</head>
<body>
  <div class="poster">
    <img class="bg" src="generated_images/layer0.png" alt="Corporate marketing team meeting in a modern office"/>
    <img class="texture" src="generated_images/layer1.png" alt="Navy tone overlay texture"/>

    <section class="content" aria-labelledby="poster-title">
      <h1 class="title" id="poster-title">TEAM</h1>
      <p class="subtitle">TOGETHER<br>EVERYONE<br>ACHIEVES MORE</p>
    </section>

    <a class="url" href="https://www.tenaha.org" aria-label="Visit www.tenaha.org">www.tenaha.org</a>
  </div>
</body>
</html>
\end{codebox}

\section{Limitations}
While DesignAsCode significantly advances the state-of-the-art in editable design generation, its current version opens up several promising avenues for future research:

\textbf{Toward End-to-End Asset Co-generation.} Our framework currently adopts a modular approach where visual assets and structural code are synthesized in a ``Plan-then-Implement" sequence. While the Visual-Aware Reflection module effectively harmonizes these components, a more tightly-coupled co-generation paradigm---where the model hallucinates pixel textures and CSS parameters simultaneously—could not only further harmonize visual styles but also significantly boost overall inference speed. This remains an intriguing frontier for multimodal research.

\textbf{Scaling to Ultra-High Information Density.} As designs scale to extreme complexity (e.g., multi-page corporate brochures or massive data visualizations), the HTML token length naturally increases. Exploring hierarchical design compression or specialized ``design tokens" presents a valuable research topic to further optimize inference efficiency for extremely intricate layouts.

\textbf{Interactive Human-in-the-Loop Co-creation.} Currently, the system is optimized for autonomous synthesis. A natural and compelling evolution would be to incorporate real-time conversational feedback into the Reflection stage. This would transform DesignAsCode from a one-shot synthesis engine into a truly interactive AI design partner, which we consider a logical next step for this paradigm.

\end{document}